\documentclass{aa}
\usepackage{txfonts}
\usepackage{amsmath}
\usepackage{amssymb}
\usepackage{natbib}
\bibpunct{(}{)}{;}{a}{}{,}%to follow the A%A style
\usepackage[dvips]{graphicx}
\usepackage[english]{babel}
\usepackage[latin1]{inputenc}
\usepackage{subfigure}
\usepackage{lscape}
\usepackage{longtable}

\begin{document}
\title{The frequency of planets in multiple systems
       \thanks{Table 8 and Appendix A are only available in 
                electronic form at www.edpsciences.org}}

\author{M. Bonavita\inst{1,2} \and S. Desidera\inst{1}}
\offprints{S.Desidera, \email{silvano.desidera@oapd.inaf.it}}
\institute{INAF - Osservatorio Astronomico di Padova, Vicolo dell'Osservatorio 5, 
           I-35122, Padova, Italy 
           \and 
           Dipartimento di Astronomia, Università di Padova, Italy}

\date{Received / Accepted }

\abstract
 {The frequency of planets in binaries is an important issue in the field of extrasolar 
  planet studies, because of its relevance in estimating of the global planet population 
  of our Galaxy and the clues it can give to our understanding of planet formation and 
  evolution. However, only preliminary estimates are available in the literature. }
% aims 
{ We analyze and compare the frequency of planets in multiple 
  systems to the frequency of planets orbiting single stars. We also try to highlight 
  possible connections between the frequency of planets and  the 
  orbital parameters of the binaries (such as the periastron  and mass ratio.)  }
% methods
{ A literature search was performed for binaries and multiple systems among the stars of the sample 
  with uniform planet detectability defined by Fischer \& Valenti (2005), and 
  202 of the 850 stars of the sample turned out to be binaries, allowing a statistical 
  comparison of the frequency of planets in binaries and single stars and a study of 
  the run of the planet frequency as a function of the binary separation.}
% results
{We found that the global frequency of planets in the binaries of the sample is not 
 statistically different from that of planets in single stars. Even conservatively taking 
 the probable incompleteness of binary  detection in our sample into account, 
we estimate that the frequency of planets in binaries can 
 be no more than a factor of three lower than that of planets in single stars.
 There is no significant dependence of planet frequency on the binary separation, 
 except for a lower value of frequency for close binaries. However, this is probably not 
 as low as required to explain the presence of planets in close binaries 
 only as the result of modifications of the binary orbit after the planet formation.
 }
{}


\maketitle

\section{Introduction}
\label{sec:intro}

The increasing  number of extra-solar planets discovered in binary or multiple stellar 
systems \citep{2004A&A...417..353E,2006ApJ...646..523R,DB06}
suggests that planets can form and survive in a variety of stellar environments.
Determining  the frequency of planets in binaries is an important issue in the 
field of extrasolar planets studies, because of its relevance in estimating the 
global planet population of our Galaxy \citep[more than half of the solar type stars are in binary 
or multiple systems as reported in][]{1991A&A...248..485D} and because of the clues it can give to our 
understanding of planet formation and evolution.
The study of the properties of planets in binaries, as well as any difference to
those of the planets orbiting single stars, could shed light on the effects 
from the presence of the companions.

A recent study by \citet{DB06} shows that the mass distribution of short 
period planets in relatively tight binaries (separation $\le 150-200$ AU) 
is significantly different from that of planets orbiting the components of 
wide binaries and single stars. There are also other possible peculiar 
features of planets in tight binaries compared to planets orbiting 
single stars, such as a lack of long-period planets and multiple planets, 
that need confirmation. The properties of exoplanets orbiting the components 
of wide binaries are instead compatible with those of planets orbiting single 
stars, except for a possible greater abundance of high-eccentricity planets.

This result  implies that the formation and/or migration and/or dynamical 
evolution processes acting in the presence of a sufficiently close external 
perturber are modified with respect to single stars. Several scenarios can 
be devised to explain it, such as a different formation mechanism for planets
in tight binaries \citep[e.g. disk instability induced by dynamical 
perturbations, as proposed by][]{2006ApJ...641.1148B}, 
and enhanced migration and accumulation rate in the presence of a stellar companions 
\citep{2000IAUS..200P.211K}, and dynamical interactions after planet formation 
\citep{2006ApJ...652.1694P}.

However, to better understand the cause of these anomalies and the origin 
itself of  the planets in very close binaries (a challenge for current 
planet formation models) a key piece of
information is missing: the frequency of planets in binaries as a function of 
the binary separation and compared to that of planets orbiting single stars.
Determining the planet frequency in binaries is made difficult by the biases 
against binaries in most of the ongoing planet search surveys 
and by the incompleteness of binary and planet detections in these samples. 

A first step in this direction has been made 
by \citet{2006IAUS...E}, performing an adaptive-optics search for 
companions around stars with and without planets 
and without previously known stellar companions from the Coralie survey.
This guarantees a fairly homogeneous binary detectability in their sample. 
However, the small number of objects  
(and the lack of confirmation of the physical association in a few cases
at the time of the presentation of their preliminary results)
did not allow them to make clear inferences on the planet frequency and 
in particular on possible differences as a function
of the binary separations.

A  much wider sample 
\citep[850 stars vs the 110 stars studied by][]{2006IAUS...E} that might be used for a study of planet 
frequency is  the `Uniform Detectability' (hereafter UD) sample 
collected by \citet{2005ApJ...622.1102F} (hereafter FV05). 
This  sample is complete for detecting planets with radial velocity (hereafter RV) 
semi--amplitude $>30$ m/s and period $<4$ yr.
However, the binarity of stars in the UD sample has not been considered up to now.

Despite some incompleteness and biases concerning binarity, this sample can be considered 
valid to draw an independent measurement of the frequency of planets in binary stars, thanks 
to the completeness of planet detection and the large sample size.
Thus, the binarity of the stars with uniform detectability was investigated 
in this work by searching some 
stellar catalogs listing stellar companions (§ \ref{sec:search}).
The result is a sub-sample of UD binaries, separated according to their different values 
of periastron and critical semimajor axis for dynamical stability of planetary orbits 
\citep[see][]{1999AJ....117..621H} (§ \ref{sec:udbinaries}). 
In this way it has been possible to compare the values of the frequency of planets in the two 
sub-samples (single stars and binary stars) and to verify a possible dependence of the frequency 
on critical semimajor axis and periastron (§ \ref{sec:resandsel}).
The biases against binaries in the original selection sample, the completeness of the binary 
detection for stars with and without planets and their impact
on the results are discussed in § \ref{sec:complet}.
The results are discussed further in § \ref{sec:discussion}, and 
in § \ref{sec:conclusion} we summarize our conclusions and offer future perspectives.

\section{The uniform detectability sample}
\label{sec:udsample}

The uniform Detectability sample has been built by considering that, despite the detectability of 
the planets' changes from star to the next and  from a survey to the other because of the different 
time span of the observations and different levels of RV errors, we can consider it complete  for companions 
with velocity amplitudes K$>$30 m/s and orbital periods shorter than 4 years.
Then, beginning from the initial target list, which  included 1330 stars observed by the Lick, Keck, and Anglo Australian 
Surveys, FV05  selected a subsample of 850 stars that satisfy these entries provided that at least 10 observations 
spanning four years were available.
Stars that were added after a planet was discovered by other groups were not included in the sample.
However, stars independently present in one of these surveys were considered even if a 
planet was detected first by another group.
Only planets with K$>$ 30 m/s and orbital periods shorter than 4-years 
were considered for the study of planet frequency. 
This corresponds to Saturn-mass planets for the shortest periods and Jupiter-mass planets for 4 year orbits.

\subsection{Changes in th UD sample}
\label{sec:udchanges}

During the analysis made for our work, we made some changes to the 
original UD sample, such as:

\begin{itemize}
\item we excluded \object{51 Peg} because it was added to the considered 
target lists after the planet detection by \citet{1995Natur.378..355M} \citep[][Fischer 2005 
private communications]{1997ApJ...481..926M} 

\item  \object{$\tau$ Boo} and \object{$\upsilon$ And} were marked in FV05 as ``without planets'', 
but the known companions fulfill all the selection criteria for the 
UD sample (Fischer 2005, private communication), so we included 
these stars as ``with planets''; 

\item \object{HD 20782} hosts a planet detected after the compilation of the 
UD sample \citep{2006MNRAS.369..249J}, but it is coherent with 
the UD requirements, so that this star has been considered as ``with planets''. 
The lack of this planet in the original UD sample confirms the hypothesis that 
high eccentricity ($e=0.93$ in this case) acts to make detection more 
difficult, as suggested by \citet{2004MNRAS.354.1165C};

\item \object{HD 18445} is flagged as ``with planets'' in the UD sample, 
probably because of a typo. It is not listed in the tables
of stars with planets. The 
RV companion ($M\sin i = 44 M_J$)  
has a mass outside the giant planet range, and it was shown to be a 
0.18 $M_{\odot}$ star orbiting  
nearly pole-on by both astrometry \citep{2000A&A...355..581H} and 
direct imaging \citep{2004A&A...425..997B}, 
so we considered it as ``without planets''
(see Appendix \ref{app:remarks} for details);

\item we considered \object{HD 196885}  as ``without planets'' because the companion reported in FV05 was not  
confirmed by \citet{2006ApJ...646..505B} (see appendix \ref{app:remarks} for details);

\end{itemize}  

The modified UD sample that is the result of those changes was then searched for companions, 
in order to build a sample of UD binaries.

\section{Searching for UD binaries}
\label{sec:search}

In order to identify known or claimed companions for the stars included in the UD sample, we checked 
available sources listing stellar companions.
Some of the most important sources are listed below:

\begin{itemize}
\item \textit{The Hipparcos and Thyco Catalogues} \citep{1997hity.book.....P};

\item \textit{The Catalogue of the Components of Double and Multiple Stars (CCDM)}  \citep{2002yCat.1274....0D}; 

\item \textit{The Washington Visual Double Star Catalog (WDS)}\citep{1997A&AS..125..523W};  

\item \textit{Sixth Catalog of Orbits of Visual Binary Stars}\citep{wds_orbit};

\item \textit{The Catalogue of Nearby Stars, Preliminary 3rd Version (C3)} \citep{1991adc..rept.....G};

\item \citet{2004ApJS..150..455G}: \textit{New HIPPARCOS-based parallaxes for 424 faint stars};

\item \citet{2002ApJS..141..503N}: \textit{Radial Velocities for 889 late-type stars}; 

\item \citet{2000A&A...356..529A}: \textit{Wide binaries among high-velocity and metal-poor stars}; 

\item \citet{2005AJ....129.2420M}: \textit{Proper motion derivatives of binaries};

\item \citet{1997A&AS..124...75T}: \textit{MSC - a catalogue of physical multiple stars}\footnote{Updated 
      version available at www.ctio.noao.edu/\ $~$atokovin/stars/}

\item \citet{2002AJ....124.1144L} \textit{Orbits of 171 single-lined spectroscopic binaries};

\item \citet{1986BICDS..30..129H} \textit{Common proper motions stars in AGK3}; 

\item \textit{The revised NLTT catalogue} \citep{2003ApJ...582.1011S};

\item  \citet{2005ApJS..159..141V} \textit{Spectroscopic properties of cool stars. I. (VF05)};

\item \textit{ The Tycho Double Star Catalogue (TDSC)} \citep{2002A&A...384..180F};

\item \textit{A Catalog of Northern Stars with Annual Proper Motions Larger than 0.15'' (LSPM-NORTH Catalog)}"
              \citep{2007AJ....133..889L}

\item \textit{SB9: The ninth catalogue of spectroscopic binary orbits} \citep{2004A&A...424..727P}

\item \textit{2MASS All Sky Catalog of point sources} \citep{2003tmc..book.....C} has 
been used only for deriving JHK photometry (used for mass determination) and for 
common proper motion confirmation, not for a search for further companions.
\end{itemize}

\noindent We also consider additional references for individual objects 
     (see Table 8 and Appendix \ref{app:remarks}).

\subsection{Selection criteria}
\label{sec:selectioncriteria}

After this search, we excluded from our UD binary sample the stars with 
non confirmed companions and with companions listed in CCDM but with 
inconsistent proper motions (and without other indication of binarity found in literature), 
and considered those stars as singles.
Stars with long-term RV and/or astrometric trends were included in the binary
sample, on the basis of the dynamical evidence of a companion. 
The RV trends we included \citep[from][]{2002ApJS..141..503N} cause an
overall RMS of RV faster than 100 m/s that cannot be due to planetary companions.
We also included stars with brown dwarf companions in the sample of binaries.
At small separation, the existence of the brown dwarf desert \citep[see][]{2006ApJ...646..505B}
guarantees little ambiguity in the classification of an object as a
massive planet or a brown dwarf (a couple of individual
cases are listed in Appendix \ref{app:remarks}). 
At large separations, where brown dwarfs companions are probably more
frequent \citep{2001ApJ...551L.163G}, this ambiguity should be taken into account
but is again limited to a few individual cases, e.g. \object{HD 206860}, which
has a T dwarf companion of mass 
$0.021 \pm 0.09~M_{\odot}$ according to \citet{2007ApJ...654..570L}, smaller
than the mass limit for planetary companions adopted by \citet{2006ApJ...646..505B}.
The small number of brown dwarf companions makes this issue irrelevant for the global results.

\subsection{The sub-sample of UD binaries}
\label{sec:udbinaries}

The properties of the UD binaries, selected from the modified UD sample, are listed in Table 8.
The stars with both components included in the UD sample are listed twice, otherwise only the star under planet scrutiny is listed.
If more than one companion is known, we report only the closer one, because of its stronger influence on planetary formation/stability.
For hierarchical triple (or higher-order multiplicity) systems, for which the 
isolated star is included in the UD sample, we sum the mass of the closest pair to consider its effective dynamical influence.
The minimum mass is listed for single-lined spectroscopic binaries.

For each object we report: 
\begin{itemize}
\item the HD number;
\item the projected separation;
\item the eccentricity, if the binary orbit is available. Otherwise we assume $e=0.31$, which is a median value between those adopted by \citet{1992ApJ...396..178F} and \citet{1991A&A...248..485D}.
\item the semi major axis in AU. 
For those stars for which the binary orbit is not available, the semimajor axis was estimated from 
the projected separation $a(AU)=1.31*\rho$ (arcsec) $*d$ (pc) \citep{1992ApJ...396..178F,1991A&A...248..485D};
\item the mass of the objects, from \citet{2005ApJ...622.1102F} (hereafter VF05)\footnote{the exceptions are listed in Appendix \ref{app:remarks}};

\item the companion mass
\begin{description}
\item a) from VF05 if both components are included in the UD sample;
\item b) calculated with the mass-luminosity relations derived by  \citet{1997AJ....113.2246R,2000A&A...364..217D};
\item c) from other literature sources (listed in the Table caption);

\end{description}
\item the critical semi-major axis for dynamical stability of planetary companions on circular 
orbits coplanar with the binary orbit, calculated using the equation 
\begin{eqnarray}
\label{eqn:acrit}
a_{crit}=\left(0.464 - 0.380 \mu - 0.631 e_{bin} +  0.586 \mu e_{bin}\right)a_{bin}+ \nonumber \\
 \left( 0.150 e_{bin}^2 - 0.198 \mu e_{bin}^2 \right) a_{bin}
\end{eqnarray}

\noindent by  \citet{1999AJ....117..621H}. In Eq. \ref{eqn:acrit}: 
$\mu=\frac{M_{com}}{M_{obj}+M_{com}}$, $a_{bin}$ is the semi major axis and $e_{bin}$  is the eccentricity of the binary orbit.
\end{itemize}

\begin{figure}
\begin{center}
\resizebox{\hsize}{!}{\includegraphics[height=6.5cm]{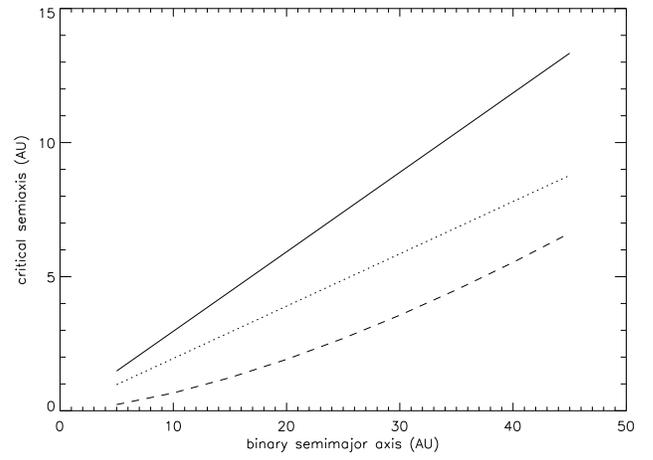}}
\end{center}
\caption{\footnotesize Values of $a_{crit}$ (solid line) and $a_{cross}$ (dashed line) and $a_{tid}$ (dotted line), versus binary semimajor axis, for e=0.31,  $M_{com}$ = 0.5 $M_{\odot}$, and $M_{obj}$=1$M_{\odot}$.}% and t=$10^4$ yr.}
\label{fig:a_crit}
\end{figure}

We chose $a_{crit}$ as a reference value, because it is a physical quantity that represents, better than 
the semi-major axis or the projected separation, the 
dynamical effects of the presence of the companion first on the circumstellar region and then on planet formation and stability.
This feature, in fact, includes both the orbital parameters and the mass ratio, and represents the maximum 
value of the semimajor axis for stable planetary orbits around the planet hosts.
The value of $a_{crit}$ is higher than the limit of the region in which the encounter velocities of planetesimal is small
enough to allow the accretion of kilometer-sized planetesimals  \citep[$a_{cross}$,][]{2006Icar..183..193T}. 
The radius of tidal truncation of the circumstellar disk $a_{tid}$ \citep{2005MNRAS.359..521P,2006ApJ...652.1694P} 
is intermediate between $a_{crit}$ and $a_{cross}$.
Figure \ref{fig:a_crit} shows the values of $a_{crit}$ (solid line), $a_{cross}$, and $a_{tid}$, 
versus binary semimajor axis, for fixed values of eccentricity and 
masses ($e=0.3$,  $M_{com}$ =0.5 $M_{\odot}$, and $M_{obj}$=1$M_{\odot}$).

%\begin{equation}\label{eqn:a_tid}
%a_{tid}=\left[0.733 f_E\left(q\right)\right]\left(1-e_{bin}\right)^{1.2}a_{bin}\\
%\end{equation}
%Where $\mu$ is defined as in Eq. \ref{eqn:acrit}, q is the mass-ratio $q=\frac{M_{comp}}{M_{obj}}$ 
%and $f_E$ = $\left(q\right)=0.49\left[0.6+q^{-2/3}\ln \left(1+q^{1/3}\right)\right]^{-1} $.}

%\begin{equation} \label{eqn:across}
%a_{cross}=0.37\frac{\left(1-e_{bin}^2\right)^{1.07}}{e_{bin}^{0.36}}\left(\frac{m_{com}}{1M_{\odot}}\right)^{-0.39}\left(\frac{a_{bin}}{10AU}\right)^{1.53}\left(\frac{t}{10^4 yr}\right)^{-0.36}AU
%\end{equation}

\section{Results and selection effects}
\label{sec:resandsel}
%\subsection{Results}
\label{sec:results}

The result of this search for companions of the UD stars is a sub-sample of 202 objects, 
15 of those having planets, so the global frequency of planets in the UD binary sample is 7.4\%.
The frequency of planets in the UD single stars sub-sample is 5.3\% (see Table 1). %\ref{tab:abin}).
The two frequencies are compatible within their errors\footnote{The errors 
reported in Tables 1 and 2 are calculated with the equation
%\ref{tab:abin} and \ref{tab:peri}
\begin{equation*}
\sigma_f= \left(N_{planets}^{-1/2} + N_{star}^{-1/2}\right)*\left(\frac{N_{planets}}{N_{stars}}\right)
\end{equation*} 

\noindent They do not include the additional error due to the incompleteness of
binary detection.}, and the slightly higher value of the global frequency in the binary sub-sample 
is probably due to some incompleteness in the sample, which is discussed in Sect.~\ref{sec:complet}.

The rather large sample size allows us to divide stars in some sub-samples according to different values of 
critical semimajor axis for dynamical stability of planets (hereafter $a_{crit}$; see Eq. \ref{eqn:acrit}).
All the stars with RV and/or astrometric trend and without direct imaging 
identification or full orbit characterization (37 objects)
were included in the closest bin, as it is likely that the companion responsible for 
the trend has a small separation.

\begin{figure}
\begin{center}
%\resizebox{\hsize}{!}{\includegraphics[height=6.5cm]{UDtest/fig2.ps}}
\resizebox{\hsize}{!}{\includegraphics[height=6.5cm]{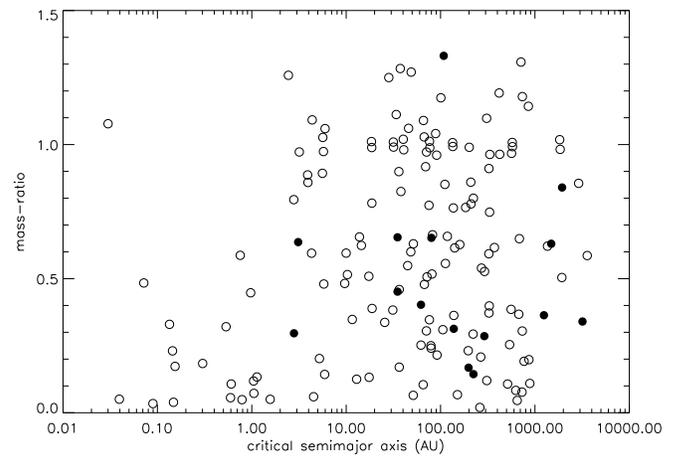}}

\end{center}
\caption{\footnotesize Critical semimajor axis \citep[from][]{1999AJ....117..621H} vs mass-ratio for 
          the binaries with planets (filled circles) and without planets (open circles) in the UD sample.}
\label{fig:a_bin}
\end{figure}

\begin{figure}
\begin{center}
%\resizebox{\hsize}{!}{\includegraphics[height=6.5cm]{UDtest/fig3.ps}}
\resizebox{\hsize}{!}{\includegraphics[height=6.5cm]{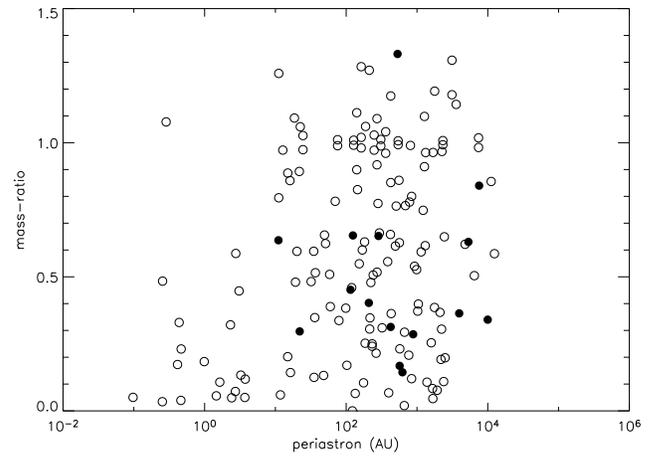}}

\end{center}
\caption{\footnotesize Periastron vs mass-ratio for the binaries with 
          planets (filled circles) and without planets (open circles) in the UD sample.}
\label{fig:p_bin}
\end{figure}

We also binned the UD binary sample according with the periastron because this allowed us to make a
direct comparison with theoretical expectations, such as those of \citet{2006ApJ...652.1694P}.
The resulting values of the frequency are listed in Tables 1 and 2, together with 
the characteristics of each sample.
%\ref{tab:abin} and \ref{tab:peri}
In Table 1  we also show the values of frequency for the complete UD binary sample %\ref{tab:abin}
and for the UD single sub-sample.
Figures \ref{fig:a_bin}-\ref{fig:p_bin} show $a_{crit}$ and periastron vs mass ratios for the
binaries of the sample with and without planets.

\begin{figure}
\subfigure[]{\includegraphics[height=6.5cm]{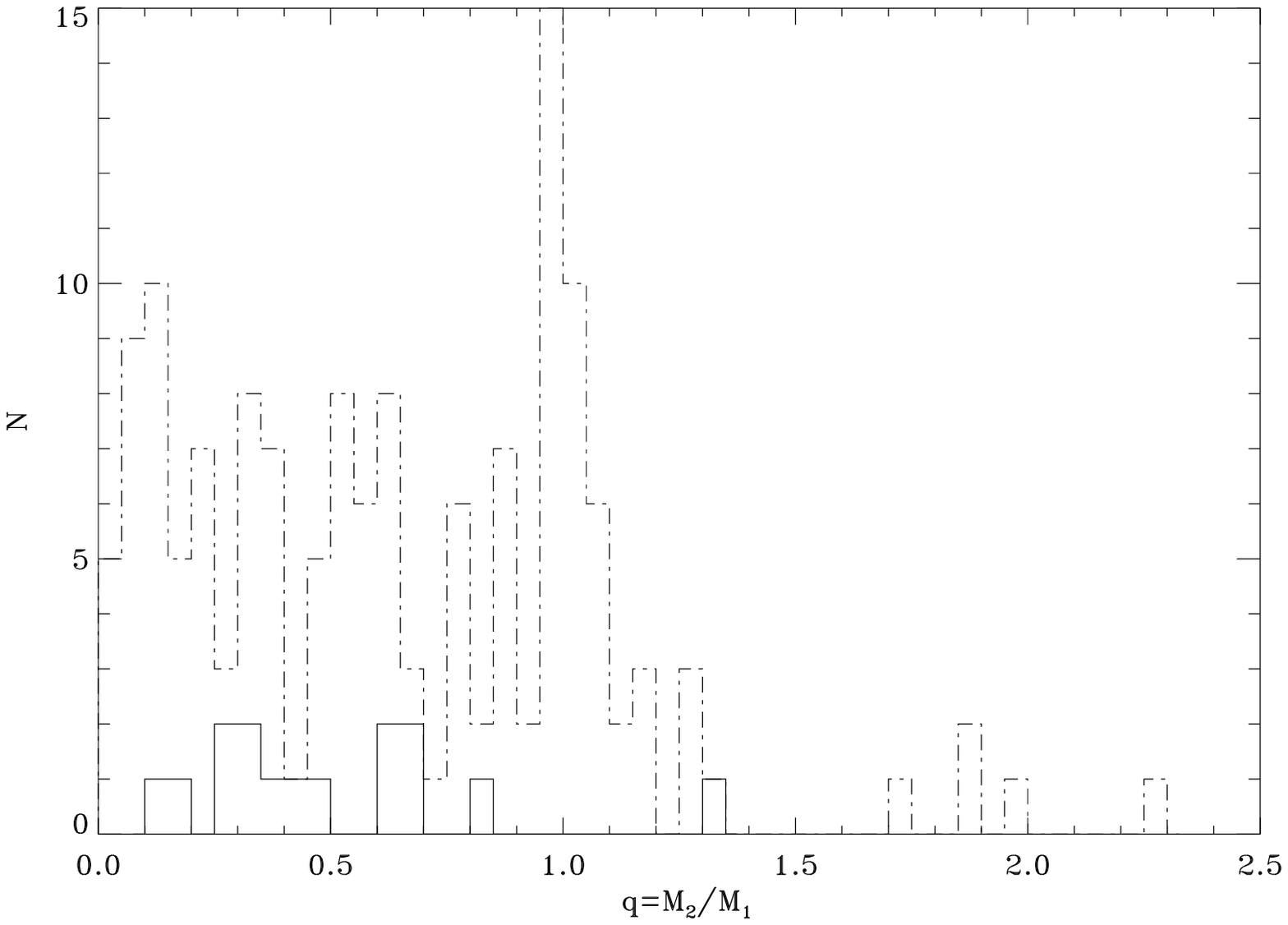}}\qquad\qquad
\subfigure[]{\includegraphics[height=6.5cm]{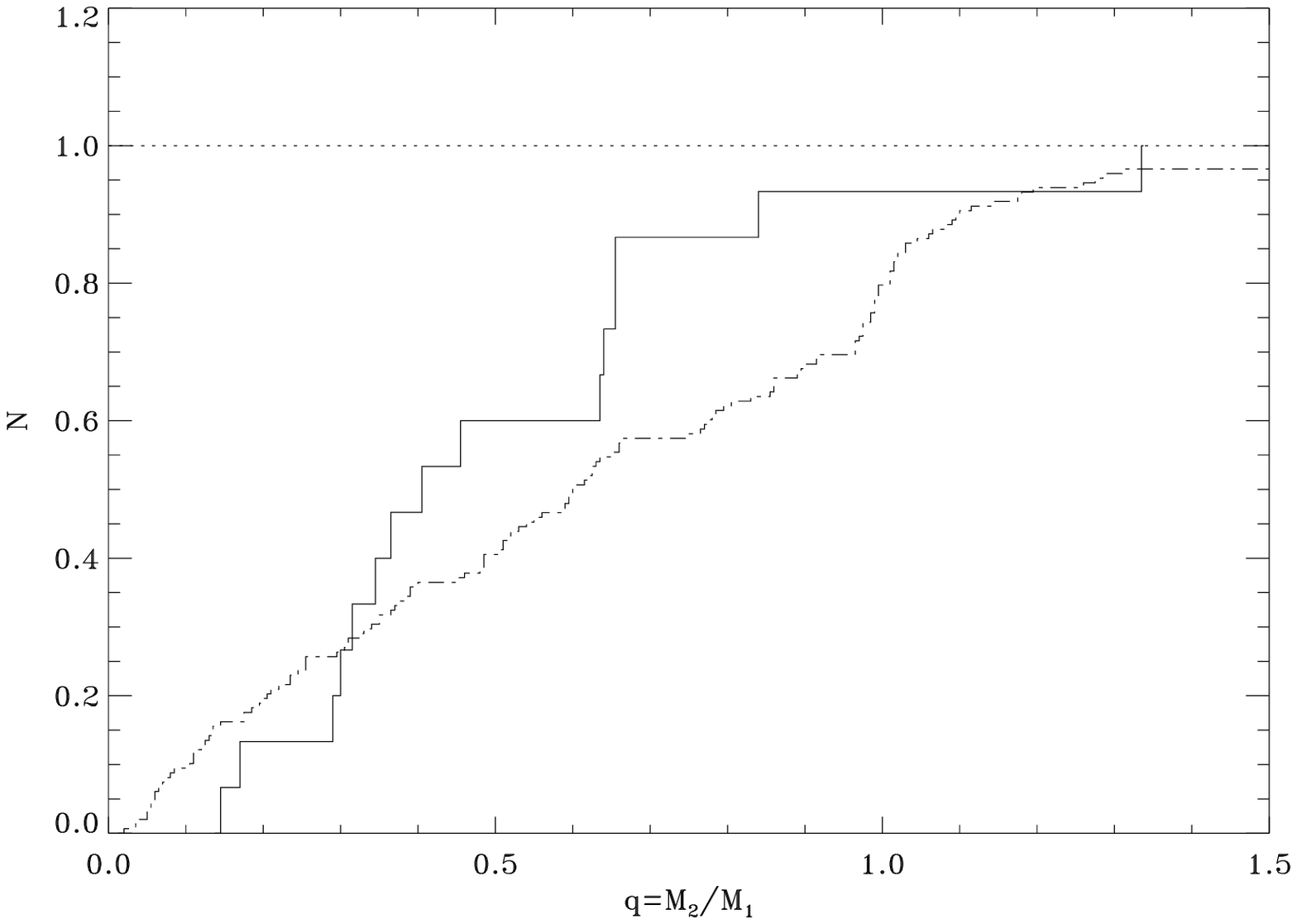}}\qquad\qquad
\caption{\footnotesize Distribution of mass-ratio values for star with (solid line) and without (dashed line) planets.
Panel (a) shows the histogram and panel (b) the cumulative distribution.}
\label{fig:q_hist}
\end{figure}

%Figure \ref{fig:a_bin} and \ref{fig:p_bin} 
These figures suggest that the binary components hosting planets usually have low-mass secondaries, 
for both low and  high separation values.
This item is confirmed by Fig. \ref{fig:q_hist} which shows the histogram of the mass-ratio values for the 
two populations.
In order to better investigate the dependence of the planet frequency on the binary mass-ratio, 
we performed a Kolmogorov Smirnov Test (hereafter KST) on the two populations.
The resulting value for the KST probability is $\sim11\%$ for the entire sample (excluding objects with 
trends, for which the value of the secondary mass is unknown) and $\sim 10\%$ for the stars 
with $a_{crit}$ $>$ 20 AU.
The difference between the  distributions is not significant enough to allow us to confirm or 
reject the hypothesis that the presence of planets is favored in binaries with a low value of the mass ratio.
This feature, as discussed in Sect. \ref{s:bias2}, could also be due to how 
the stars hosting planets are  preferentially searched for low-mass 
companions, and this is probably one of the causes of their lower observed values of the mass ratio.
On the other hand, the lack of planet detections up to now in the ongoing RV planet search using SARG
at TNG \citep{2006tafp.conf..119D}, which targets only binaries with similar components 
and with a typical separation between 100 to 400 AU,
suggests that the mass ratio might be an important parameter for the occurrence of planets.
Therefore, the possible role of the mass ratio on the frequency of planets remains an open point
that requires further investigations.

\begin{table}
\begin{center}
\begin{tabular}{|c|c|c|c|}
\hline
  $a_{crit}$                          & $N_{star}$ & $N_{planets}$ & $\frac{N_{planets}}{N_{stars}}$\\
\hline
                    $<$  20 AU& 89&  2&0.022$\pm$0.018\\
                    20 - 50 AU& 18&  2&0.111$\pm$0.105\\
                   50 - 100 AU& 24&  2&0.083$\pm$0.076\\
                  100 - 250 AU& 26&  4&0.154$\pm$0.107\\
                    $>$ 250 AU& 45&  5&0.111$\pm$0.066\\
\hline
\hline
         UD Singles sub-sample&647& 34&0.053$\pm$0.011\\
   Entire UD binary sub-sample&202& 15&0.074$\pm$0.024\\
\hline
\end{tabular}
\caption{\footnotesize Frequency of planets in binaries with different values of  $a_{crit}$.}
\end{center}
\label{tab:abin}
\end{table}

\begin{table}
\begin{center}
\begin{tabular}{|c|c|c|c|}
\hline
 Periastron         & $N_{star}$ & $N_{planets}$ & $\frac{N_{planets}}{N_{stars}}$\\
\hline
         $<$   50 AU& 81&  2&0.025$\pm$0.020\\
         50 - 200 AU& 28&  2&0.071$\pm$0.064\\
        200 - 500 AU& 30&  3&0.100$\pm$0.076\\
       500 - 1000 AU& 23&  4&0.174$\pm$0.123\\
         $>$ 1000 AU& 39&  4&0.103$\pm$0.068\\
\hline
\end{tabular}
\caption{Frequency of planets in binaries with different values of  periastron.} %\footnotesize
\end{center}
\label{tab:peri}
\end{table}

\subsection{Completeness and selection effects}
\label{sec:complet}

These results suggest that the planet frequency in binaries and
single stars is similar. 
However, selection effects concerning binaries in the original definitions of the
Lick, Keck, and AAT samples  and possible incompleteness of binary detection
might alter the results; we discuss them here in order to better estimate the frequency
of planets in binaries.

\subsubsection{Estimate of the missing binaries}
\label{sec:inputbias}

The first issue to consider is the completeness of the binary census in the sample.
We can derive an upper limit to the incompleteness by assuming that the stars in the 
UD sample have the same binary frequency as in \citet{1991A&A...248..485D} (57\%).
The number of missing binaries in the sample would then be $\sim 296$ \footnote{For the 
estimate of the number of unrecognized binaries, we excluded (and then consider 
them as single stars) the stars with brown dwarfs companions, because the substellar companions are not 
included in the statistical analysis by \citet{1991A&A...248..485D}; hence, the number of binaries found in the 
UD sample decreases to 188.}.

To derive a lower limit to the frequency of planets in binaries,
we assume that all the ``missing" binaries are without planets.
We then obtain $f_{bin}=3.09 \pm 0.94 \%$ and $f_{sin}=9.31 \pm 0.69 \%$; i.e. the frequency
of planets in binaries cannot be less than one third of that of planets orbiting single stars.
We stress that this is a very conservative lower limit  on the frequency of planets in binaries (an upper
limit on the incompleteness of the binary detection in the sample),
because the UD sample has  selection biases against binarity.
In fact the input target lists exclude spectroscopic binaries  and binaries with separations 
less than 2 arcsec known at the time of the target selection 
\citep[see][]{2002MNRAS.337.1170J, 2005PThPS.158...24M, 2004ApJS..152..261W}. 
Therefore we expect that the absolute binary frequency in the UD sample is significantly lower than
the unbiased one derived by \citet{1991A&A...248..485D}; the distribution of the orbital parameters 
and mass ratio of the binaries in the UD sample should also be different with respect to 
unbiased samples.

To take this selection bias in account, we did a Monte Carlo simulation 
making an estimate of the fraction of binaries with $\rho < $2" expected on the 
basis of the period and mass-ratio distribution assumed by \citet{1991A&A...248..485D}
\footnote{The distribution in orbital period assumed by \citet{1991A&A...248..485D} is 
\begin{equation*}
%  f\left(\log P\right)= C \exp \left{ \frac{-\left(\log P - \overline{\log P}\right)^2}{2\sigma^2_{\log P}}\right}
f  (\log P)= C \exp  \frac{-\left(\log P - \overline{\log P}\right)^2}{2\sigma^2_{\log P}}  
\end{equation*}
\noindent where $\overline{\log P}=4.8$, $\sigma_{\log P}=2.3$ and P is in days.
For the mass-ratio $q=M_2/M_1$ they assume 
\begin{equation*}
\xi\left(q\right)=k\exp\left{\frac{-\left(q-\mu\right)^2}{2\sigma^2_q}\right}
\end{equation*}
\noindent where $\mu=2.3$, $\sigma_q=0.42$ and $k=18$ for their sample of G-Dwarfs.}.
In this way we found that $\sim 45.7 \%$ of the binaries predicted by \citet{1991A&A...248..485D} have $\rho < 2"$
for the distance distribution of the stars of the UD sample,
so we expect that a significant fraction of the `missing' binaries are given by stars excluded 
at the time of the target selection because of their small separation.

Combining these features, we find that the total number of binaries expected in the UD sample  with a separation $>2$ arcsec 
is 221. Our census yields 138 binaries (120 without considering wide companions orbiting close pairs with a separation $<2$ arcsec)
so there should be 83 missing binaries (95 without considering wide companions in triple systems) with separations $>2$ arcsec. 

These numbers are expected to hold if there are no biases besides the exclusion 
of binaries with separations smaller than 2 arcsec  in the UD sample.
However, it is possible that the stars that were added later to the samples for specific reasons
\citep[e.g. high metallicity,][]{2000ApJ...545..504B,2003ApJ...587..423T}
have different selection criteria concerning binarity.
In some individual cases, the inclusion in the input target lists  of 
certain types of systems is favored:  a few binary systems with
similar components were probably included in the sample 
after dedicated studies of chemical abundances differences between the 
components \citep{2001A&A...377..123G,2002ApJ...579..437M}.
Unfortunately, we do not have enough information about the details of the building of the sample. 
We do expect that this only plays a minor role in the global binary statistics.

\subsubsection{Completeness of the binarity of planet hosts vs non planet hosts}
\label{s:bias2}

The role of completeness in binary detection would be minor for our purposes if
the completeness itself were independent of the occurrence of planets.
Unfortunately, this is not the case.
Planet hosts are systematically searched for companions after planet discoveries.
Therefore, the completeness of the binarity of planet hosts 
is certainly greater than that of stars without planets.
This bias could cause a spurious increase in planet frequency in binaries
and should be carefully addressed.
We proceeded as follows.
We considered the binarity of planet hosts in the UD sample (including those 
with $K<30$ m/s and/or $P>4$ yr; 21 objects overall). If the binarity was found
on the basis of dedicated studies after planet discoveries, the star was
classified as single; if instead the binarity was known
independently of the planet discovery (e.g. inclusion in WDS, CCDM, Hipparcos, detection
of astrometric and RV trends, etc.), the star was kept as a binary.
This results in a change in binary status for 7 stars (4 with UD planets and 3 with non-UD planets).
In this way we obtained a binary sample that is less complete than the original one
but without biases favoring the binarity of planet hosts.
The revised  frequency of planets in binaries then becomes 11/195=0.056 and the corresponding
for single stars 38/654=0.058 (the results for all the $a_{crit}$ bins are reported in Table 3). 
The difference with respect to the full sample indicates that indeed the dedicated searches for
companions around planet hosts somewhat alters the results but the frequencies of planets in binaries
and in single stars remain similar.

\begin{table}
\begin{center}
\begin{tabular}{|c|c|c|c|}
\hline
  $a_{crit}$                          & $N_{star}$ & $N_{planets}$ & $\frac{N_{planets}}{N_{stars}}$\\
\hline
                     $<$ 20 AU& 89&  2&0.022$\pm$0.018\\
                    20 - 50 AU& 18&  2&0.111$\pm$0.105\\
                   50 - 100 AU& 22&  2&0.091$\pm$0.083\\
                  100 - 250 AU& 23&  2&0.087$\pm$0.079\\
                    $>$ 250 AU& 43&  3&0.070$\pm$0.051\\
\hline
\hline
         UD Singles sub-sample&654& 38&0.058$\pm$0.011\\
   Entire UD binary sub-sample&195& 11&0.056$\pm$0.021\\
\hline
\end{tabular}
\caption{\footnotesize Frequency of planets in binaries with different values of  $a_{crit}$, without considering
as a binary the planet-host whose companions were discovered thanks to dedicated
follow up after planet detection.}
\end{center}
\label{tab:abin2}
\end{table}

\subsubsection{Dependence on separation}

The completeness of binarity in this sample is probably a function of the separation.
At small separations, the inclusion of stars with RV and an astrometric trend 
probably allows a fairly high completeness level.
In fact, most of the binaries recently discovered by means of deep adaptive optics imaging 
(e.g. \object{HD 13445}; \object{HD 161797}, \object{HD 190406}, \object{HD 196885}; see App. \ref{app:remarks} for references) would have been included as 
binaries in this study thanks to their dynamical signatures, even without the direct imaging identification. 
Pairs with a small magnitude difference in separation between 0.2 to 10 arcsec should have been detected
by Hipparcos \citep{2000A&A...361..770Q}. Wide binaries ($\rho$ $>$ 5 - 10 arcsec) are more easily discovered and then 
included in CCDM and WDS even for larger magnitude differences.
Intermediate values of separation (e.g. 1 to 5 arcsec) are probably the most incomplete ones, 
as the detection of a low-mass companion requires dedicated high-resolution imaging that is
not available for all the stars, and the companions 
are not expected to produce detectable RV or astrometric signatures.

\subsubsection{Effects of incomplete information on orbit and masses of the companions}

The determination of $a_{crit}$ requires the availability of the full binary orbit and an estimate of the mass
the companion.
All these quantities are unknown for the 37 stars with astrometric and/or RV trends .
In any case, we included these objects in the binary statistic  because of the significance high level of the 
trends as resulting from the original works by \citet{2005AJ....129.2420M} and \citet{2002ApJS..141..503N}.
Furthermore, several objects present both the astrometric and RV signatures, which further confirms 
of their nature as multiple objects.
From the timescales of the detected orbital motion, we can reasonably infer a period short enough to allow
the inclusion of these stars in our closest bin ($a_{crit}<20$~AU).
However, any more detail on the separation distribution of these companions cannot be determined, which 
representing a major limitation for studying of the details run of planet frequency
vs $a_{crit}$ at small separations.
 
For the stars where only the projected separation is available, we have considered an approximate values for 
the semimajor axis and eccentricity values from the adopted statistical relations.  
Even if for individual objects this approximation could be quite different from the 
real value, we expect that statistically it would give a realistic representation of the true semimajor axis
and periastron distributions.

\subsection{The volume-limited sample}
\label{sec:volume}

As a further investigation of the role of the biases mentioned in Sect.~\ref{sec:complet} 
in our results, we have selected a volume-limited sample  (hereafter VLUD), including the
stars within a distance of 18 pc from the Sun, analogous with FV05.
The selection effects considered in Sect.~\ref{sec:complet} are expected to be smaller for the
VLUD sample.
In fact, the census of companions around stars in this sample is expected to be more complete than that of 
the global sample because the closest stars were in general more carefully searched for stellar companions,
and at close distance the 2 arcsec limit corresponds to a smaller
physical separation (36 AU at 18 pc). The small distance also favors  overlap between 
different detection techniques.
Indeed, only 1 star with astrometric trend and without direct imaging
identification is included in the VLUD sample.

The volume-limited UD sample  includes  129 stars, of which 44 are known binaries.
The fraction of stars with planets in the VLUD sub-sample  is 9.1\% for binaries
and 9.4\% for single stars.
We selected some bins in $a_{crit}$ and periastron also for the VLUD sample but, because of the 
small number of stars included in this sample, we considered only 2 ranges of values.
Tables 4 and 5 show the frequency values obtained for the VLUD sub-samples.  %\ref{tab:a_vlim} and \ref{tab:peri_vlim}
Statistical error bars are larger for the VLUD sample because of the small number of objects
but the frequency of planets in single stars and in binaries is again similar.

\begin{table}
\begin{center}
\begin{tabular}{|c|c|c|c|}
\hline
  $a_{crit}$                          & $N_{star}$ & $N_{planets}$ & $\frac{N_{planets}}{N_{stars}}$\\
\hline
                     $<$ 20 AU& 21&  2&0.095$\pm$0.088\\
                     $>$ 20 AU& 23&  2&0.087$\pm$0.079\\
\hline
\hline
       VLUD Singles sub-sample& 85&  8&0.094$\pm$0.043\\
 Entire VLUD binary sub-sample& 44&  4&0.091$\pm$0.059\\
\hline
\end{tabular}
\caption{\footnotesize Frequency of planets in binaries with different values 
of  $a_{crit}$ for the volume-limited sample.}
\end{center}
\label{tab:a_vlim}
\end{table}

\begin{table}
\begin{center}
\begin{tabular}{|c|c|c|c|}
\hline
 Periastron         & $N_{star}$ & $N_{planets}$ & $\frac{N_{planets}}{N_{stars}}$\\
\hline
           $<$ 50 AU& 19&  2&0.105$\pm$0.098\\
           $>$ 50 AU& 25&  2&0.080$\pm$0.072\\
\hline
\end{tabular}
\caption{\footnotesize Frequency of planets in binaries with different values of  periastron for the volume limited sample.}
\end{center}
\label{tab:peri_vlim}
\end{table}

\begin{figure}
\subfigure[]{\includegraphics[height=6.5cm]{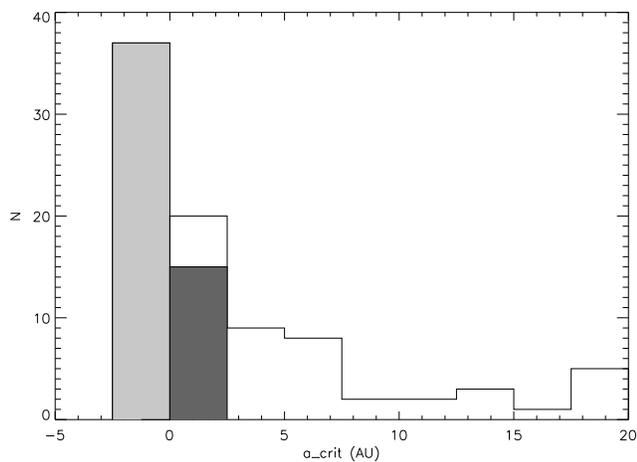}}\qquad\qquad
\subfigure[]{\includegraphics[height=6.5cm]{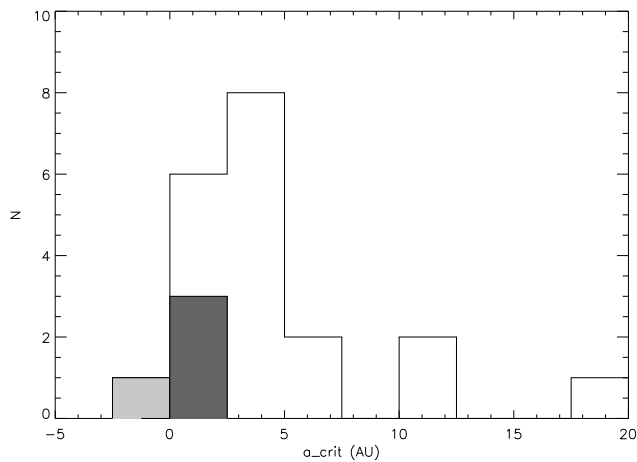}}\qquad\qquad
\caption{\footnotesize Distribution of $a_{crit}$ $<$ 20 AU for the complete UD binary sample (a) and for the VLUD sample (b).
In both panels, the light grey column contains the stars with-long term RV or astrometric trends, arbitrarily placed at $a_{crit}<0$ for display purposes. The dark grey column represents the contribute of spectroscopic binaries.}
\label{fig:hist}
\end{figure}

A possibly interesting difference with respect to the full UD sample is that 
the frequency of planets in tight binaries ($a_{crit}<20$~AU) 
is not lower than in wide binaries and single stars.
This might be explained looking at Fig. \ref{fig:hist}.
The panels show the distribution of $a_{crit}$ for the lowest bin in $a_{crit}$ for the 
complete UD sample (upper panel) and for the VLUD sample (lower panel). 
In both panels, the first column on the left contains the stars included as binaries 
only because of the long-term RV or astrometric trends (i.e. mass and separation of the companions
not known).
From these histograms we can easily see that the percentage of the stars without 
definite orbital characteristics (thus included only on the basis of dynamical signatures) is 
much larger in the complete UD sample (37/89, $\sim$ 42 \%) with respect to the VLUD sample 
in which just one object, \object{HD 120780}, is included only thanks to its astrometric trend.
At the same time, for the systems for which $a_{crit}$ can be derived, the distribution 
of the $a_{crit}$ values for the VLUD sample is 
centered on the second bin (2.5 $<$ $a_{crit}$ $<$ 5.0 AU), while in the complete sample 
there is  a relatively large number of binaries (at least 20, excluding those 
with only RV or astrometric trends) for which $a_{crit}$ is smaller than $\sim 2.5$ AU, the 
separation limit corresponding to $P$=4 yr for solar-type systems.
Therefore a lower frequency for planets in these systems is expected, as  
only a part of the separation range considered here can host planets on stable orbits.

\section{Discussion}
\label{sec:discussion}
\subsection{Global estimate of frequency of planets in binary stars.}

Using a sample made of binaries with uniform planet detectability, we compared 
the frequency of planets in binaries to that of planets orbiting single stars.
Looking at the whole binary sample we can conclude that the frequency of planets in binaries 
is not statistically different from planets orbiting single stars.
With very conservative assumptions on the incompleteness of binary detection, we 
find that the frequency of planets in the binaries of the sample cannot
be lower by more than a factor of three compared to planets orbiting single stars.
When considering  samples that are less affected by selection biases such as the VLUD 
sample, or excluding only the binaries detected thanks to dedicated follow-up after 
planet discoveries, that cause a spurious increase of the binary fraction of planet hosts, 
we found that the two frequencies are indeed very similar.
This is qualitatively consistent with the preliminary results of \citet{2006IAUS...E}.

An important point to consider is that this conclusion applies to the kind of binaries
that are included in UD sample, i.e. with a separation distribution biased against
close binaries, and cannot be directly applied to samples with unbiased binary
distribution.
Furthermore, it applies to the kind of planets fulfilling the requirements for
inclusion in the UD sample, i.e. $P<4$ yr and $K>30$~m/s.
It is possible that e.g. the effects of binarity 
on planets in wider orbits are stronger.

\subsection{Dependence on the binary separation}
\label{sec:disc_separation}
The size of our sample allowed us to divide it into some sub-groups according to the 
value of the critical semimajor axis for the dynamical stability.
In this way we can argue that there is no significant dependence of the frequency on 
$a_{crit}$ (and on the periastron) except for companions with 
$a_{crit}$ less than 20 AU (that corresponds to a separation $<$ 50-100 AU, depending on 
the mass ratio of the components).
This result makes stronger the conclusion by \citet{DB06}, who reported that the presence of 
distant companions (separation $>$ 300-500 AU) does not significantly affect the process of 
planet formation, as the mass and period distribution of planets in such wide binaries 
are similar to those of planets orbiting single stars \citep{DB06}, and 
the frequency of planets is also similar (this work).

What happens at small separations is more crucial to understanding the role of companions
in planet formation.
\citet{DB06} found that the properties of 
planets in close binaries, in particular the mass distribution, are different from those orbiting
single stars and components of wide binaries\footnote{Note that the definition of tight binaries 
used by \citet{DB06} ($a_{crit}$ $<$ 75 AU) is different than the one adopted here.}.
We showed in this study indication for a lower frequency of planets in tight binaries.

The frequency of planets in close binaries can be used to further investigate how these
planets formed and the origin of their anomalous properties.
Indeed, \citet{2006ApJ...652.1694P} shows that knowing the value of the frequency of 
planets in close binaries\footnote{Defined as those binaries with semi-major axis less 
than 50 AU} should allow two alternative formation scenarios to be distinguished. 
A low frequency (about 0.1\% but with an uncertainty of about one order of magnitude, so we 
can consider 1\% as a limit-value) 
would be compatible with dynamical interactions that cause the formation of the tight binary 
after planet formation.

\begin{table}
\begin{center}
\begin{tabular}{|c|c|c|}
\hline
Frequency & Probability & Probability \\
          & full sample & VLUD sample \\
\hline
\hline
0.1\%     & 0.3\%       & 0.02\%      \\
1\%       & 16.3\%      &  1.7\%      \\
5\%       & 11.3\%      & 19.8\%      \\
10\%      & 0.4\%       & 28.4\%      \\
\hline
\end{tabular}
\caption{\footnotesize Probability of observing 2 binaries with planets in a sample of 
89 (resp. 21) binaries in the UD (resp. VLUD) sample 
with $a_{crit}<20$ AU, for different planet frequencies.}
\end{center}
\label{tab:binomial_acrit}
\end{table}

\begin{table}
\begin{center}
\begin{tabular}{|c|c|c|}
\hline
Frequency & Probability & Probability \\
          & full sample & VLUD sample \\
\hline
\hline
0.1\%     & 0.3\%      &  0.02\%      \\
1\%       & 14.6\%      &  1.4\%      \\
5\%       & 14.1\%      & 17.8\%      \\
10\%      & 0.79\%      & 28.5\%      \\
\hline
\end{tabular}
\caption{\footnotesize Probability of observing 2 binaries with planets in a sample of 81 (resp. 19) 
binaries in the UD (resp. VLUD) sample with periastron $<50$ AU, for different planet frequencies.}
\end{center}
\label{tab:binomial_peri}
\end{table}

We tested the probability of obtaining the observed number of close binaries with planets
in the UD and VLUD samples for different frequencies of planets using the binomial
distribution.
This leads to being able to confidently exclude (99\%, see Tables 6 \& 7) 
the preferred value in \citet{2006ApJ...652.1694P}.
The observed frequency is marginally compatible only with 
the upper limit on planet frequency by \citet{2006ApJ...652.1694P} ($f \sim 1\%$)
for the UD sample, and hardly compatible for the VLUD sample.
The nominal probabilities derived here should be taken with some caution because
of the possible incompleteness in binary detection at small separations
(anyway estimated to be small, in particular for the VLUD sample, see above)
and because the  separation distribution is different than the
unbiased samples.
Nevertheless, 
our relatively high frequency of planets in close binaries  suggests that the dynamical
interaction after planet formation is not the unique channel for creating this kind of systems.
Instead, we can infer that  giant planets might form in binaries that have a small separation at the time 
of planet formation, possibly in a different way from planets around single stars 
(and around components of wide binaries).

%Finally, if we consider a volume limited sample, the results seems to be different  for the lowest $a_{crit}$ bin.
%In fact, the frequency of planets in binaries with $a_{crit}$ less than 20 AU appears higher 
%for the stars in the VLUD sample with respect to those in the complete UD sample.

The run of planet frequency at small separations might shed more light
on the formation mechanism(s) of planets in binaries.
\citet{DB06} noted a possible paucity of planets in binaries with a critical semimajor axis 
for dynamical stability in the range 10 - 30 AU;
only one planet was found in this range, while there are 5 planets with $a_{crit}$ less than 
10 AU and 4 planets with 30 $<$ $a_{crit}$ $<$ 50 AU.
A bimodal distribution of planet frequency, with 
a secondary maximum at $a_{crit} \sim 3-5$  AU, is suggested by these
data, and it
would explain the different characteristics of planets in tight binaries
as the result of a different formation mechanism.
However, our work is not able to confirm or reject the reality of such a feature,
as only 1 star in the UD sample has 10 $<$ $a_{crit}$ $<$ 20 AU and
11 have 20 $<$ $a_{crit}$ $<$ 30 AU (none of them with planets), making the lack
of planet detections insignificant.
Furthermore, the closest bin ($a_{crit}< 20$ AU or periastron $< 50$ AU) includes
several stars (37 out of 89) for which a direct detection of the companion is missing and which 
are included only on the basis of the astrometric and/or RV trends. 
Without a determination of the physical parameters of these companions,
the study of the run of planet frequency at small separations is not possible.

%Both the  confirmation of the binarity of these stars (with a determination of the 
%physical parameters of these companions) 
%and the completion of on-going surveys focused on binaries 
%\citep{2006tafp.conf..119D,2006IAUS...E,2005ApJ...626..431K} would be useful 
%to explain in more details the run of frequency of planets at small separation.

%\subsection{Dependence on  mass-ratio}

\section{Conclusions}
\label{sec:conclusion}

After a detailed search for binarity for all stars in the UD sample 
collected by \citet{2005ApJ...622.1102F}, we compared the frequency of planets 
in binaries and single stars. It turns out that the two frequencies of planets 
are fairly similar.
Even taking possible incompleteness in the binary 
detection into account in a very conservative way, the frequency of planets in the binaries
of the sample cannot be more than a factor of three lower than that of planets orbiting 
single stars.   

For moderately wide binaries, the frequency of planets is independent on separation.
Considering the similar mass and period distributions of planets 
orbiting single stars and components of wide binaries,
we then concluded that a wide companion plays a marginal role in the formation and evolution of giant planets. 

On the other hand, we found a lower frequency of planets in close binaries
than in single stars and components of 
wide binaries. However, this is  probably not as low as required to explain 
the occurrence of planets in close binaries  only as the result of modifications in
the binary orbit after planet formation.
This, together with the differences in the properties of planets in tight binaries
\citep{DB06}, suggests that planets  do form in tight binaries in spite 
of the  unfavorable conditions, 
possibly in a different way 
than for planets around single stars. 
However, crucial issues still need clarification.
In fact, 
it is not yet clear if the run of the planet frequency when moving to smaller separations
is characterized by a continuous decrease, by a sharp cut off at 
which the differences on planet frequency  characteristics suddenly onset,
or by a bimodal distribution, with a minimum between 10 $<$ $a_{crit}$ $<$ 30 AU,
a relative maximum at $a_{crit} \sim 3-5$  AU, a further decrease to zero at
extremely small separations, as the zone for dynamical stability of planets vanishes.

These open points might be clarified by a detailed characterization of the binaries
in current samples of RV surveys 
(completeness of binary detection and, when possible, full determination of the
orbital elements) and by  the completion of dedicated surveys searching for planets in binaries
\citep{2006tafp.conf..119D,2006IAUS...E,2005ApJ...626..431K}.
%The availability of a larger and more complete sample will allow us to better understand the behavior 
%of the planet frequency in binaries and, at the same time, to disentangle the questions about the formation 
%of planets in these peculiar environments and especially  about the formation mechanisms and the different 
%characteristics of the planets in tight binaries. 

\begin{acknowledgements}

   This research has made use of the 
   SIMBAD database, operated at the CDS, Strasbourg, France,
   of the Washington Double Star Catalog maintained at the U.S. Naval Observatory,
   and of data products from the Two Micron All Sky Survey.
    We warmly thank D.~Fischer for kindly providing information
    on the UD sample and S.~Lepine for providing the complete tables
    of his work before publication.
    We thank the anonymous referee for comments that allowed us
    to improve the content and the presentation of the paper.
   We thank R.~Gratton for his comments and suggestions.
   This work was funded by COFIN 2004 
   ``From stars to planets: accretion, disk evolution and
   planet formation'' by the Ministero Univ. e Ricerca
   Scientifica Italy and by  PRIN 2006
   ``From disk to planetary systems: understanding the origin
  and demographics of solar and extrasolar planetary systems'' 
   by INAF.

\end{acknowledgements}

%\nocite{*}

\bibliographystyle{aa}
\bibliography{bib_paper}

\Online

\longtab{8}{
\begin{longtable}{|l|c|c|c|c|c|c|c|c|}
\caption{\label{tab:binary_UD} \footnotesize   Properties of binaries found in the UD sample: 
projected separation (arcsec), eccentricity and semi major-axis (when available), 
masses of the object and  the companion, and critical semimajor axis for dynamical stability 
of planets \citep{1999AJ....117..621H}. 
For systems for which only the projected separation was available (empty spaces in eccentricity column) 
the semi major axis was derived from the projected separation using the 
relation $a$(AU)=1.31$\rho$(arcsec)$d$(pc) \citep[see][]{2002PASP..114..529F,1991A&A...248..485D}.
The asterisk in the last column marks systems discussed individually in Appendix \ref{app:remarks}.
The mass flag indicates the source for the companion mass: 
\textbf{a:} $M_{comp}$ from VF06;  \textbf{b:} $M_{comp}$ from 
\citet{1997AJ....113.2246R,2000A&A...364..217D}; 
\textbf{ c:} $M_{comp}$ from individual papers (see Reference below).}
\endfirsthead
\hline
\multicolumn{9}{l}{\footnotesize \textit{(continued on next page)}}\\
\endfoot
\multicolumn{9}{l}{\textit{(Table 8 - continued from previous page)}}\\
\hline
HD      &  $\rho$  & ecc      & a        &$a_{crit}$& $M_{obj}$&$M_{com}$ & Mass Flag & Remarks                       \\
        &(arcsec)  &          & (AU)     &   (AU)   &($M_{\odot}$)&($M_{\odot}$)&         &                           \\
\hline
\endhead
\multicolumn{9}{l}{}\\
\caption*{\footnotesize \textbf{Remarks:} \textbf{P:} Stars with planets as in FV05;  
\textbf{S:} Stars without planets as in FV05;  
\textbf{SB:} Spectroscopic Binaries; 
\textbf{RV:} Stars with RV linear trends \citep[see][]{2002ApJS..141..503N}; 
\textbf{ ${\Delta\mu}$: } Stars with discrepant proper motion in Hipparcos and Thyco II \citep[see][]{2005AJ....129.2420M};  
\textbf{G:} Stars with accelerating proper motions in Hipparcos. \citep[see][]{2005AJ....129.2420M}.}
\multicolumn{9}{l}{}\\
\caption*{\footnotesize \textbf{References}: 
\textbf{\object{HD 3074:}} \citet{2000A&A...356..529A}; 
\textbf{\object{HD 4614:}} \citet{1983PUSNO..24g...1W}; 
\textbf{\object{HD 4747:}} \citet{2002ApJS..141..503N}; 
\textbf{\object{HD 7693:}} \citet{2004A&A...418..989N}; 
\textbf{\object{HD 10360} - \object{HD 10361}:} \citet{1983PUSNO..24g...1W}; 
\textbf{\object{HD 11964:}} \citet{2000A&A...356..529A}; 
\textbf{\object{HD 13445:}} \citet{2006A&A...459..955L,DB06}; 
\textbf{\object{HD 13507:}} \citet{2003A&A...410.1039P};
\textbf{\object{HD 13531:}} \citet{metchev}; % S. Metchev, PhD Thesis;
\textbf{\object{HD 13612:}} \citet{1967AJ.....72..899W,1991A&A...248..485D};
\textbf{\object{HD 16141:}} \citet{DB06,2004A&A...417.1031M}; 
\textbf{\object{HD 16160:}} \citet{2000A&A...356..529A}, \citet{2000AJ....120.2082G}; 
\textbf{\object{HD 16895:}} \citet{1983PUSNO..24g...1W}; 
\textbf{\object{HD 18445:}} \citet{1991A&A...248..485D,2000A&A...355..581H,2001ApJ...562..549Z}; 
\textbf{\object{HD 20782:}} \citet{DB06}; 
\textbf{\object{HD 23439:}} \citet{2000A&A...356..529A}; 
\textbf{\object{HD 27442:}} \citet{2006A&A...456.1165C,DB06}; 
\textbf{\object{HD 29836:}} \citet{1981AJ.....86..588G};
\textbf{\object{HD 30649:}} \citet{2002ApJS..141..503N,2002yCat.1274....0D}; 
\textbf{\object{HD 31412:}} \citet{2002ApJS..141..503N,1991adc..rept.....G}; 
\textbf{\object{HD 35956:}} \citet{2002ApJ...568..352V};
\textbf{\object{HD 38529:}} \citet{2006A&A...449..699R, DB06}; 
\textbf{\object{HD 39587:}} \citet{2002ApJS..141..503N}; 
\textbf{\object{HD 43587:}} \citet{2002ApJ...568..352V,1991A&A...248..485D,2003ApJ...582.1011S}; 
\textbf{\object{HD 64468:} }\citet{2002ApJ...568..352V};
\textbf{\object{HD 65430:}} \citet{2002ApJS..141..503N}; 
\textbf{\object{HD 65907:}} \citet{1997A&AS..124...75T};
\textbf{\object{HD 72760:}} \citet{metchev};  %%S. Metchev, PhD Thesis;
\textbf{\object{HD 77407:}} \citet{2004A&A...417.1031M}, \citet{metchev}; % S. Metchev, PhD Thesis;
\textbf{\object{HD 86728:}} \citet{2000MNRAS.311..385G,2007AJ....133..889L};
\textbf{\object{HD 90839:}} \citet{1991adc..rept.....G,1991A&A...248..485D}; 
\textbf{\object{HIP 52940:}} \citet{2002ApJS..141..503N}; 
\textbf{\object{HD 92222:}} \citet{2002A&A...384..180F}, this paper;
\textbf{\object{HD 97334:}} \citet{2005AJ....129.2849B};
\textbf{\object{HD 101177:}} \citet{1991A&A...248..485D,2003ApJ...582.1011S};
\textbf{\object{HD 111398:}} \citet{2007AJ....133..889L};
\textbf{\object{HD 120066:}} \citet{2000A&A...356..529A,2004ApJS..150..455G};
\textbf{\object{HD 120136:}} \citet{2006ApJ...646..523R,DB06}; 
\textbf{\object{HD 120237:}} \citet{2000A&A...356..529A}; 
\textbf{\object{HD 120780:}} \citet{2005AJ....129.2420M};
\textbf{\object{HD 122742:}} \citet{2002ApJS..141..503N};  
\textbf{\object{HD 128620} - \object{HD 128627}:}\citet{1983PUSNO..24g...1W}; 
\textbf{\object{HD 131156:}}  \citet{1983PUSNO..24g...1W,1991A&A...248..485D}; 
\textbf{\object{HD 131511:}} \citet{2002ApJS..141..503N}; 
\textbf{\object{HD 134440/39:}}\citet{2000A&A...356..529A}; 
\textbf{\object{HD 135101:}} \citet{2004A&A...420..683D}; 
\textbf{\object{HD 139323:}} \citet{1983PUSNO..24g...1W}; 
\textbf{\object{HD 139477:}} \citet{2007AJ....133..889L};
\textbf{\object{HD 140913:}} \citet{2002ApJS..141..503N}; 
\textbf{\object{HD 146362 B:}} \citet{1983PUSNO..24g...1W,1997A&AS..124...75T};
\textbf{\object{HD 150554:}} \citet{metchev};     %S. Metchev, PhD Thesis;
\textbf{\object{HD 156274:}} \citet{1983PUSNO..24g...1W}; 
\textbf{\object{HD 161797:}} \citet{1983PUSNO..24g...1W,2002ApJS..141..503N,2006AJ....132..177W,2005AJ....129.2420M}; 
\textbf{\object{HD 167215:}} \citet{2005AJ....129.2420M};
\textbf{\object{HD 169822:}} \citet{2002ApJ...568..352V};
\textbf{\object{HD 174457:}} \citet{2002ApJS..141..503N};
\textbf{\object{HD 178911 B:}} \citet{2000AstL...26..116T};
\textbf{\object{HD 184860:}} \citet{2002ApJ...568..352V};
\textbf{\object{HD 185395:}} \citet{2007AJ....133..889L};
\textbf{\object{HD 187691:}} \citet{1991A&A...248..485D};
\textbf{\object{HD 190360:}} \citet{2000A&A...356..529A, DB06}; 
\textbf{\object{HD 190406:}} \citet{2002ApJ...571..519L};
\textbf{\object{HD 191408:}} \citet{2000A&A...356..529A}; 
\textbf{\object{HD 195019:}} \citet{2000A&A...356..529A, DB06}; 
\textbf{\object{HD 197076:}} \citet{1991A&A...248..485D};
\textbf{\object{HD 196885:}} \citet{2006A&A...456.1165C};
\textbf{\object{HD 198387:}} \citet{2005AJ....129.2420M};
\textbf{\object{HD 206860:}} \citet{2006astro.ph..9464L}
\textbf{\object{HD 208776:}} \citet{2002ApJS..141..503N}; 
\textbf{\object{HD 213519:}} \citet{2007AJ....133..889L};
\textbf{\object{HD 219542:}} \citet{2004A&A...420..683D}; 
\textbf{\object{HD 219834:}} \citet{1997A&AS..124...75T};
\textbf{\object{HD 221830:}} \citet{2000A&A...356..529A}.} 
\endlastfoot
\hline
HD       &  $\rho$  & ecc      & a        &$a_{crit}$& $M_{obj}$&$M_{com}$ & Mass Flag  & Remarks                      \\
         &(arcsec)  &          & (AU)     &   (AU)   &($M_{\odot}$)&($M_{\odot}$)&         &                           \\
\hline
   531 A &     5.30 &          &   482.30 &    76.37 &     1.64 &     1.66 &           a&                             S\\
   531 B &     5.30 &          &   482.30 &    76.94 &     1.66 &     1.64 &           a&                             S\\
   3074  &     4.80 &          &   225.89 &    38.01 &     1.20 &     0.99 &           b&                             S\\
   3651  &    43.20 &          &   623.38 &   151.00 &     0.89 &     0.06 &           b&                        S (*) \\
   3770  &          &          &          &          &     1.25 &          &            &                        S, RV \\
   3795  &          &          &          &          &     1.94 &          &            &            S, RV, $\Delta\mu$\\
   3821  &     8.45 &          &   284.51 &    51.50 &     1.00 &     0.63 &           b&                             S\\
   4614  &    12.49 &     0.49 &    72.00 &    10.30 &     0.99 &     0.51 &           b&                             S\\
   4747  &          &     0.64 &     6.70 &     0.79 &     0.82 &     0.04 &           c&                        S, SB \\
   6734  &          &          &          &          &     1.08 &          &            &               S, $\Delta\mu$ \\
  6872 A &    14.60 &          &  1442.48 &   271.29 &     1.91 &     1.03 &           b&                             S\\
  6872 B &    14.60 &          &  1442.48 &   187.22 &     1.03 &     1.91 &           a&                             S\\
   7693  &     0.90 &     0.04 &    23.40 &     5.98 &     0.84 &     0.89 &           b&                        S (*) \\
   8765  &          &          &          &          &     1.20 &          &            &            S, $\Delta\mu$, G \\
  10360  &    11.20 &     0.53 &    52.20 &     5.66 &     0.75 &     0.77 &           a&                             S\\
  10361  &    11.20 &     0.53 &    52.20 &     5.74 &     0.77 &     0.75 &           a&                             S\\
  11964  &    40.50 &          &  1790.10 &   325.00 &     1.13 &     0.67 &           b&                        S (*) \\
  13043  &    79.20 &          &  3809.52 &   684.35 &     1.14 &     0.74 &           b&                             S\\
  13507  &          &     0.14 &     4.30 &     1.57 &     1.00 &     0.05 &           c&       S, SB, $\Delta\mu$ (*) \\
  13531  &     0.70 &          &    23.30 &     5.21 &     0.94 &     0.19 &           c&                             S\\
13612 B  &    16.70 &          &  1000.83 &   121.14 &     1.02 &     2.32 &           c&                        S (*) \\
  16141  &     6.20 &          &   289.35 &    62.00 &     1.15 &     0.29 &           b&                         S (*)\\
  16160  &     3.30 &     0.75 &    15.00 &     1.04 &     0.76 &     0.09 &           c&                        S (*) \\
  17037  &          &          &          &          &     1.23 &          &            &                        S, RV \\
  16895  &    20.50 &     0.13 &   249.50 &    76.20 &     1.24 &     0.43 &           b&                             S\\
  18143  &    44.10 &          &  1312.86 &   222.85 &     0.90 &     0.72 &           b&                          S(*)\\
  18445  &     0.10 &     0.56 &     1.06 &     0.14 &     0.78 &     0.18 &           c&                        S (*) \\
  20766  &   310.00 &          &  4876.30 &   711.69 &     0.91 &     1.19 &           a&                             S\\
  20807  &   310.00 &          &  4876.30 &   736.00 &     0.95 &     1.12 &           a&                             S\\
  21019  &     3.90 &          &   187.08 &    36.48 &     1.11 &     0.51 &           b&                             S\\
  23439  &     8.00 &          &   254.80 &    37.41 &     0.67 &     0.86 &           c&                        S (*) \\
  26491  &          &          &          &          &     1.00 &          &            &               S, $\Delta\mu$ \\
 28255 A &     5.90 &          &   199.42 &    31.78 &     1.07 &     1.06 &           a&                             S\\
 28255 B &     5.90 &          &   199.42 &    31.60 &     1.06 &     1.07 &           a&                             S\\
  29461  &          &          &          &          &     1.20 &          &            &                        S, RV \\
  29836  &   100.00 &          &  5590.00 &   852.14 &     1.19 &     1.36 &           c&                        S (*) \\
  30339  &          &     0.25 &     0.13 &     0.04 &     1.39 &     0.07 &           c&                        S, SB \\
  30649  &          &          &          &          &     0.90 &          &            &                    S, RV (*) \\
  31412  &          &          &          &          &     1.17 &          &            &                    S, RV (*) \\
  32923  &     0.18 &     0.90 &     2.86 &     0.03 &     1.03 &     1.11 &           b&                             S\\
  33473  &   100.00 &          &  7475.00 &  1358.02 &     1.32 &     0.82 &           b&                             S\\
  35956  &          &     0.62 &     2.60 &     0.30 &     0.98 &     0.18 &           c&                     S, SB (*)\\             
  37394  &    97.50 &          &  1546.35 &   292.41 &     0.93 &     0.49 &           b&                             S\\
  39587  &          &     0.45 &     5.90 &     1.13 &     1.05 &     0.14 &           c&                        S, SB \\
  40397  &     4.10 &          &   123.66 &    25.62 &     0.92 &     0.31 &           b&                             S\\
  43587  &          &     0.80 &    11.60 &     0.53 &     1.06 &     0.34 &           c&                    S, SB (*) \\
  44120  &    40.40 &          &  1911.73 &   330.66 &     1.23 &     0.92 &           b&                             S\\
  45701  &          &          &          &          &     1.18 &          &            &               S, $\Delta\mu$ \\
  45588  &    41.20 &          &  1601.44 &   325.89 &     1.21 &     0.45 &           b&                             S\\
  47157  &    10.10 &          &   502.88 &   105.72 &     1.13 &     0.35 &           b&                             S\\
  50281  &    58.30 &          &   659.37 &   118.04 &     0.76 &     0.50 &           b&                             S\\
  50639  &          &          &          &          &     1.16 &          &            &            S, RV, $\Delta\mu$\\
  51929  &          &          &          &          &     0.86 &          &            &               S, $\Delta\mu$ \\
  53705  &    21.00 &          &   425.88 &    69.47 &     0.97 &     0.89 &           a&                        S (*) \\
  53706  &    21.00 &          &   425.88 &    65.90 &     0.89 &     0.97 &           a&                        S (*) \\
  61606  &    58.10 &          &  1072.53 &   184.33 &     0.81 &     0.62 &           b&                             S\\
  63754  &     5.60 &          &   376.38 &    71.81 &     1.50 &     0.76 &           b&                             S\\
  64468  &          &     0.26 &     0.56 &     0.15 &     0.81 &     0.14 &           c&                        S, SB \\       
  65907  &    60.00 &          &  1263.60 &   201.45 &     0.99 &     0.98 &           c&                        S (*) \\
  65277  &     4.10 &          &    93.27 &    18.82 &     0.72 &     0.28 &           b&                S, $\Delta\mu$\\
  65430  &          &     0.32 &     4.00 &     1.05 &     0.83 &     0.06 &           c&                        S, SB \\
  66171  &    49.00 &          &  3013.01 &   730.27 &     0.91 &     0.07 &           b&                             S\\
  72760  &     0.90 &          &    25.51 &     5.92 &     0.91 &     0.13 &           c&            S, $\Delta\mu$ (*)\\   
  72780  &          &          &          &          &     1.28 &          &            &                        S, RV \\
  73668  &    26.50 &          &  1229.86 &   210.35 &     1.13 &     0.88 &           b&                             S\\
  77407  &     1.60 &          &    50.00 &     9.65 &     1.12 &     0.54 &           c&                        S (*) \\
  86728  &   134.00 &          &  2595.58 &   626.32 &     1.08 &     0.09 &           b&                        S (*) \\
  88218  &     2.00 &          &    79.82 &    14.49 &     1.09 &     0.68 &           b&                             S\\
  90839  &   122.80 &          &  2043.39 &   372.01 &     1.12 &     0.69 &           b&                        S (*) \\
  92222  &    17.70 &          &  2070.90 &   332.90 &     1.09 &     1.05 &           c&                        S (*) \\
  92987  &          &          &          &          &     1.15 &          &            &               S, $\Delta\mu$ \\
HI 52942 &    17.60 &          &  2779.92 &   418.07 &     1.04 &     1.24 &           b&                        S (*) \\
HI 52940 &          &     0.37 &     2.60 &     0.60 &     1.12 &     0.12 &           c&                     S, SB (*)\\
  97334  &    90.00 &          &  2620.00 &   649.08 &     1.09 &     0.05 &           c&                        S (*) \\
  99491  &    28.80 &          &   666.43 &   111.12 &     1.01 &     0.86 &           a&                             S\\
  99492  &    28.80 &          &   666.43 &   100.71 &     0.86 &     1.01 &           a&                        S (*) \\
 100180  &    15.40 &          &   460.46 &    82.24 &     1.10 &     0.73 &           b&                             S\\
 100623  &    17.00 &          &   209.95 &    51.31 &     0.77 &     0.05 &           b&                             S\\
 101177  &     9.70 &          &   293.81 &    45.85 &     0.99 &     1.05 &           c&                        S (*) \\
 102365  &    23.00 &          &   275.08 &    65.42 &     0.86 &     0.09 &           b&                             S\\
 103432  &    73.20 &          &  3520.92 &   565.26 &     0.92 &     0.89 &           b&                             S\\
 103829  &          &          &          &          &     1.20 &          &            &                        S, RV \\
104556 B &          &          &          &          &     1.12 &          &            &            S, $\Delta\mu$, G \\
 105113  &     6.60 &          &   440.15 &    75.43 &     1.28 &     0.99 &           b&                             S\\
 107705  &    20.00 &          &   774.80 &   141.13 &     1.22 &     0.75 &           b&                             S\\
 111031  &          &          &          &          &     1.14 &          &            &               S, $\Delta\mu$ \\
 111398  &          &          &    75.00 &    17.52 &     1.06 &     0.14 &           b&                             S\\
111484 A &     8.80 &          &   846.56 &   134.25 &     1.38 &     1.39 &           a&                             S\\
111484 B &     8.80 &          &   846.56 &   134.84 &     1.39 &     1.38 &           a&                             S\\
 114729  &     8.00 &          &   364.00 &    79.00 &     1.00 &     0.25 &           b&                         S (*)\\
 116442  &    26.50 &          &   564.98 &    90.90 &     0.76 &     0.73 &           a&                             S\\
 116443  &    26.50 &          &   564.98 &    88.68 &     0.73 &     0.76 &           a&                             S\\
 120066  &   488.50 &          & 19432.53 &  3581.91 &     1.16 &     0.68 &           b&                             S\\
 120237  &    11.60 &          &   426.76 &    81.04 &     1.16 &     0.60 &           b&                             S\\
 120476  &     3.40 &     0.44 &    33.15 &     4.36 &     0.76 &     0.83 &           b&                             S\\
 120690  &          &          &          &          &     1.02 &          &            &               S, $\Delta\mu$ \\
 120780  &          &          &          &          &     0.74 &          &            &         S, G, $\Delta\mu$ (*)\\
 121384  &    33.00 &          &  1634.49 &   328.24 &     1.18 &     0.47 &           b&                             S\\
 122742  &          &     0.48 &     5.30 &     0.75 &     0.92 &     0.54 &           c&                        S, SB \\
 125455  &    15.30 &          &   413.71 &    91.76 &     0.79 &     0.17 &           b&                             S\\
 126614  &    41.90 &          &  3725.75 &   883.39 &     1.19 &     0.13 &           b&                             S\\
 128428  &     0.80 &          &    54.39 &     9.99 &     1.26 &     0.75 &           b&                             S\\
 128621  &    17.51 &     0.51 &    22.76 &     2.44 &     0.89 &     1.12 &           a&                        S (*) \\
 128620  &    17.51 &     0.51 &    22.76 &     2.79 &     1.12 &     0.89 &           a&                        S (*) \\
 128674  &   490.00 &          & 17453.80 &  2906.33 &     0.83 &     0.71 &           b&                             S\\
 129814  &          &          &          &          &     1.06 &          &            &            S, RV, $\Delta\mu$\\
 131156  &     4.90 &     0.51 &    32.80 &     3.93 &     0.92 &     0.79 &           b&                        S (*) \\
 131511  &          &     0.51 &     0.52 &     0.07 &     0.93 &     0.45 &           c&                        S, SB \\
 131977  &    24.90 &          &   190.98 &    28.29 &     0.76 &     0.95 &           c&                        S (*) \\
 131923  &          &          &          &          &     1.04 &          &            &            S, $\Delta\mu$, G \\
 133161  &          &          &          &          &     1.18 &          &            &               S, $\Delta\mu$ \\
 134440  &   302.00 &          & 11581.70 &  1830.52 &     0.55 &     0.56 &           a&                             S\\
 134439  &   302.00 &          & 11581.70 &  1850.83 &     0.56 &     0.55 &           a&                             S\\
 134331  &    50.60 &          &  1993.13 &   325.83 &     1.12 &     1.02 &           a&                             S\\
 134330  &    50.60 &          &  1993.13 &   307.70 &     1.02 &     1.12 &           a&                             S\\
 135101  &    23.50 &          &   870.67 &   209.00 &     1.07 &     0.92 &           c&                             S\\
 136580  &          &          &          &          &     1.17 &          &            &            S, RV, $\Delta\mu$\\
 137778  &    51.90 &          &  1403.38 &   182.11 &     0.90 &     1.67 &           c&                        S (*) \\
 139323  &   121.90 &          &  3533.88 &   468.58 &     0.89 &     1.55 &           c&                        S (*) \\
 139477  &    42.00 &          &  1042.86 &   221.21 &     0.75 &     0.22 &           b&                             S\\
 140913  &          &     0.54 &     0.55 &     0.09 &     1.17 &     0.04 &           c&                        S, SB \\
 140901  &     8.10 &          &   160.06 &    36.49 &     1.00 &     0.17 &           b&                             S\\
 142229  &          &          &          &          &     1.09 &          &            &                        S, RV \\
 144579  &    70.10 &          &  1312.27 &   308.93 &     0.75 &     0.09 &           b&                             S\\
 145435  &          &          &          &          &     1.19 &          &            &               S, $\Delta\mu$ \\
145958 A &     4.20 &     0.39 &   124.00 &    18.66 &     0.90 &     0.89 &           a&                        S (*) \\
145958 B &     4.20 &     0.39 &   124.00 &    18.53 &     0.89 &     0.90 &           a&                             S\\
 146362  &     6.80 &     0.76 &   130.00 &     5.03 &     1.12 &     2.19 &           c&                        S (*) \\
 147722  &     5.40 &          &   220.43 &    33.89 &     1.16 &     1.29 &           a&                             S\\
 147723  &     5.40 &          &   220.43 &    36.17 &     1.29 &     1.16 &           a&                             S\\
 149806  &     5.90 &          &   154.17 &    31.20 &     0.94 &     0.36 &           b&                             S\\
 150554  &    11.60 &          &   675.58 &   138.12 &     1.13 &     0.41 &           c&                             S\\
 150248  &          &          &          &          &     0.96 &          &            &               S, $\Delta\mu$ \\
 151090  &   163.60 &          & 10102.30 &  1929.65 &     1.17 &     0.59 &           b&                             S\\
 156274  &     8.65 &     0.78 &    91.65 &     4.30 &     0.79 &     0.47 &           b&                             S\\
 157466  &          &          &          &          &     0.92 &          &            &               S, $\Delta\mu$ \\
 159909  &          &          &          &          &     1.04 &          &           b&                             S\\
 161797  &     1.42 &     0.32 &    22.00 &     3.90 &     1.15 &     0.13 &           c&         S, RV, $\Delta\mu$ (*) \\
 164595  &    88.00 &          &  3306.16 &   674.36 &     0.98 &     0.36 &           b&                             S\\
 166553  &     1.40 &          &    77.17 &    13.83 &     1.22 &     0.80 &           b&                             S\\
 167215  &          &          &          &          &     1.15 &          &            &         S, G, $\Delta\mu$ (*)\\
 167665  &          &          &          &          &     1.11 &          &            &                        S, RV \\
 169586  &          &          &          &          &     1.29 &          &            &               S, $\Delta\mu$ \\
 169822  &          &     0.48 &     0.84 &     0.13 &     0.91 &     0.30 &           c&                     S, SB (*)\\       
 173667  &    48.20 &          &  1196.81 &   266.63 &     1.54 &     0.32 &           b&                             S\\
 174457  &          &     0.23 &     1.90 &     0.60 &     1.07 &     0.06 &           c&                        S, SB \\
 175345  &     5.40 &          &   348.89 &    67.43 &     1.17 &     0.56 &           b&                             S\\
 179957  &     8.20 &          &   254.77 &    40.25 &     1.01 &     1.03 &           a&                             S\\
 179958  &     8.20 &          &   254.77 &    40.73 &     1.03 &     1.01 &           a&                             S\\
 179140  &     0.50 &          &    34.06 &     5.60 &     1.12 &     1.00 &           b&                             S\\
 184860  &          &     0.67 &     1.40 &     0.15 &     0.77 &     0.03 &           c&                     S, SB (*)\\ 
 185395  &    37.00 &          &   894.66 &   196.56 &     1.34 &     0.31 &           b&                             S\\
 187691  &    14.40 &          &   363.17 &    79.35 &     1.37 &     0.33 &           b&                        S (*) \\
 190360  &   188.60 &          &  3898.36 &   864.00 &     1.01 &     0.20 &           b&                         S (*)\\
 190406  &     0.80 &          &    18.41 &     4.52 &     1.09 &     0.065&           c&                    S, RV (*) \\
 190067  &     2.86 &          &    55.00 &    12.91 &     0.80 &     0.10 &           b&                          S(*)\\
 190771  &          &          &          &          &     1.07 &          &            &            S, RV, $\Delta\mu$\\
 191785  &   103.80 &          &  2766.27 &   559.07 &     0.83 &     0.32 &           b&                             S\\
 191408  &     7.10 &          &    56.30 &    11.60 &     0.69 &     0.24 &           b&                             S\\
 192343  &    43.40 &          &  3627.81 &   577.95 &     1.28 &     1.27 &           b&                             S\\
 192344  &    43.40 &          &  3627.81 &   575.18 &     1.27 &     1.28 &           b&                             S\\
194766 B &    43.60 &          &  2635.62 &   423.63 &     1.10 &     1.06 &           b&                             S\\
 195564  &     2.90 &          &    91.23 &    17.39 &     1.12 &     0.57 &           b&                             S\\
 196201  &     2.20 &          &   109.54 &    18.72 &     0.87 &     0.68 &           b&                             S\\
 197076  &   125.00 &          &  3412.50 &   767.57 &     0.99 &     0.19 &           b&                        S (*) \\
 196068  &    17.40 &          &   882.18 &   159.88 &     1.69 &     1.06 &           b&                             S\\
 196885  &     0.70 &          &    30.03 &     5.80 &     1.25 &     0.60 &           c&                       S, (*) \\
 198387  &          &          &          &          &     1.32 &          &            &         S, G, $\Delta\mu$ (*)\\
 199598  &          &          &          &          &     1.15 &          &            &               S, $\Delta\mu$ \\
 200565  &          &          &          &          &     1.06 &          &            &            S, RV, $\Delta\mu$\\
 206387  &     3.70 &          &   264.07 &    48.40 &     1.20 &     0.72 &           b&                             S\\
 206860  &    43.20 &          &  1033.34 &   260.95 &     1.07 &     0.021&           c&                             S\\
 208776  &          &     0.27 &     4.20 &     0.97 &     1.14 &     0.51 &           c&                        S, SB \\
 212330  &    81.10 &          &  2161.31 &   513.18 &     1.12 &     0.12 &           b&                             S\\
 212168  &    20.10 &          &   603.60 &   112.67 &     1.06 &     0.59 &           b&                             S\\
 213519  &    62.00 &          &  3481.92 &   734.00 &     1.05 &     0.32 &           b&                             S\\
 214953  &     7.80 &          &   239.30 &    44.82 &     1.13 &     0.62 &           b&                             S\\
 215578  &          &          &          &          &     1.02 &          &            &                     S, RV (*)\\
 215648  &    11.80 &          &  2485.08 &   538.88 &     1.26 &     0.32 &           b&                             S\\
 217004  &     8.90 &          &   794.86 &   136.69 &     1.27 &     0.97 &           b&                             S\\
 218101  &           &          &          &          &     1.26 &          &            &               S, $\Delta\mu$ \\
219542 A &     5.28 &          &   388.00 &    71.00 &     1.08 &     1.05 &           c&                          S(*)\\
219542 B &     5.28 &          &   388.00 &    67.00 &     1.05 &     1.08 &           c&                          S(*)\\
 219834  &    13.00 &          &   331.24 &    48.80 &     0.74 &     0.94 &           c&                        S (*) \\
 220077  &     0.20 &          &    19.89 &     3.19 &     1.09 &     1.06 &           b&                             S\\
 221830  &     8.00 &          &   335.92 &    70.79 &     0.95 &     0.29 &           b&                             S\\
 223084  &          &          &          &          &     1.09 &          &            &            S, RV, $\Delta\mu$\\
\hline
\multicolumn{9}{l}{} \\
\multicolumn{9}{l}{} \\
\multicolumn{9}{l}{\footnotesize Table 8: Stars with planets}\\
\hline
HD       &  $\rho$  & ecc      & a        &$a_{crit}$& $M_{obj}$&$M_{com}$ & Mass Flag  & Remarks                      \\
         &(arcsec)  &          & (AU)     &   (AU)   &($M_{\odot}$)&($M_{\odot}$)&      &                              \\
\hline
    142  &     5.40 &          &   179.71 &    35.00 &     1.24 &     0.56 &           b&                             P\\
   9826  &    55.50 &          &   974.02 &   223.00 &     1.32 &     0.19 &           b&                             P\\
  13445  &     1.30 &     0.40 &    18.40 &     3.10 &     0.77 &     0.49 &           c&           P, $\Delta\mu$ (*) \\
  20782  &   252.20 &          & 11802.96 &  1940.00 &     1.00 &     0.84 &           c&                        P (*) \\
  27442  &    13.80 &          &   326.51 &    62.00 &     1.49 &     0.60 &           c&                        P (*) \\
  38529  &   283.02 &          & 15600.10 &  3190.00 &     1.47 &     0.50 &           b&                     P, G (*) \\
  40979  &   192.20 &          &  8320.34 &  1488.00 &     1.19 &     0.75 &           b&                             P\\
  46375  &    10.30 &          &   447.23 &    80.00 &     0.92 &     0.60 &           b&                             P\\
  75732  &    84.90 &          &  1379.63 &   291.00 &     0.91 &     0.26 &           b&                             P\\
 120136  &    12.00 &     0.91 &   245.00 &     2.80 &     1.35 &     0.40 &           b&                        P (*) \\
178911 B &    13.60 &          &   830.96 &   108.00 &     1.42 &     1.89 &           c&                        P (*) \\
 188015  &    13.00 &          &   888.94 &   198.00 &     1.25 &     0.21 &           b&                             P\\
 195019  &     4.00 &          &   194.48 &    35.00 &     1.07 &     0.70 &           b&                             P\\
 196050  &    10.90 &          &   664.57 &   138.00 &     1.15 &     0.36 &           b&                             P\\
 222582  &   113.30 &          &  6171.45 &  1246.00 &     0.99 &     0.36 &           b&                             P\\
\hline
\end{longtable}
}

\newpage

\begin{appendix}

\section{Comments on individual objects}
\label{app:remarks}
\subsection{List of included binaries}
\label{app:inclusions}
\begin{itemize}

\item \object{HD 3651}: planet outside the UD limits. The companion is a cool brown dwarf. 

\item \object{HD 7693}:
Hipparcos lists a companion at 0.9 arcsec ($\sim 19$ AU). Furthermore, the star is listed in CCDM as the wide 
companion of HD 7788, with a separation of 319 arcsec ($\sim 6731$ AU).
The values of parallaxes and proper motion in right ascension would suggest a physical bounding, and 
the RV values reported by \citet{2004A&A...418..989N} seems to confirm this hypothesis. 
But the RV values for HD 7788 has large errors, and the proper motion in declination reported by 
Hipparcos are in disagreement.
Probably this peculiarity could be explained by considering that both HD 7693 and HD 7788 are 
close binaries themselves.
In fact Hipparcos lists a companion at 0.9 arcsec for the first star, and one at 5 arcsec = 105 AU, for the 
last one.
Because we are interested in the effect of the binarity on the planetary 
formation/evolution, we will take only the closest companion to HD 7693 into account .

\item \object{HD 11964}: planet outside the UD limits.

\item \object{HD 13445} (GL 86): 
the companion was discovered by \citet{2001A&A...370L...1E} and it was 
classified as a brown dwarf, but successive work shows that the secondary is a 
$\sim$ 0.5 $M_{\odot}$ white dwarf \citep{2005MNRAS.361L..15M,2006A&A...459..955L}.
\citet{DB06} described a possible evolution of the system during the mass-loss phase
of the originally more massive star.
The white dwarf companion is responsible for the observed RV and astrometric trends.

\item \object{HD 13507}:
suspected for some time of harboring a planet, but the later measurements 
obtained with ELODIE invalidated this interpretation and instead revealed 
a classical spectroscopic binary velocity curve, caused by a 
low-mass ($m\sin i \sim 50 M_J$) companion \citep{2003A&A...410.1039P}.
Adaptive optics imaging did not directly detect the companion.

\item \object{HD 13612}: 
triple system. AB is a CPM pair. A is an SB2 \citep{1991A&A...248..485D} 
and the disagreement between spectral and photometric parallaxes of A and 
B are probably due to its nature. An additional component at 2'.9 
is optical \citep{1967oeds.conf..221W};
the component included in the UD sample is \object{HD 13612 B} and, because the mass 
of this star is not listed in VF05, we derive it according  
to the \citet{1997AJ....113.2246R} and \citet{2000A&A...364..217D} mass-luminosity 
calibration.

\item \object{HD 16141}: planet outside the UD limits.

\item \object{HD 16160}=GL105: triple system composed of an inner pair 
and another wide companion ($\rho=164.8$ arcsec = 1200 AU, mass $0.38~M_{\odot}$).
The orbit of the close pair was derived by \citet{2000AJ....120.2082G}
($a=15$ AU $e=0.75$). The mass of the close companion is about $0.09~M_{\odot}$.

\item \object{HD 18143}: triple system.
Component B at 6.5 arcsec = 149 AU, component C at 43 arcsec = 985 AU.

\item \object{HD 18445}: is a member of a quintuple system (component C).
Components AB (HD 18455) form a visual binary with 
P=147 yr and a=1.55 arcsec = 40 AU \citep{1983PUSNO..24g...1W},
%\citet{1991A&A...248..485D};
C is at 27.2 arcsec $\sim 700$ AU and is a spectroscopic binary with minimum mass
in the brown dwarf range  \citep[$m \sin i = 0.042~M_{\odot}$ according to][]{2001ApJ...562..549Z}. 
\citet{2000A&A...355..581H} demonstrated 
that the pair is close to face-on and derived a mass of 0.176 $M_{\odot}$.
The close companion was also visually resolved by \citet{2004A&A...425..997B}  at 0.1 arcsec = 2.6 AU.
A further CPM companion (D) is at a projected separation 5.090 arcsec = 130 AU from C.

\item \object{HD 20782}: listed as a binary in the CCDM catalog. 
 The companion is HD 20781. \citet{DB06} suggest a physical association 
 and consider HD 20782 and HD 20781 as a very wide CPM pair.

\item \object{HD 23439}: triple system. The secondary is a spectroscopic binary
(period = 49 days, masses 0.74+0.12 $M_{\odot}$).

\item \object{HD 27442} ($\epsilon$ Ret): 
a companion at $\sim$13 arcsec $\sim 237$ AU is included in WDS and was confirmed by 
\citet{2006A&A...456.1165C}.  
\object{HD 27442 B} is probably  a white dwarf  with a mass of about 
0.6 $M_{\odot}$ \citep[see][]{DB06}.

\item \object{HD 29836}: triple system. \object{HD 285970} is a wide companion 
of the star included in the UD sample. \object{HD 285970} is also
a short-period spectroscopic binary \citep{1981AJ.....86..588G}.
Another companion to \object{HD 29836} at 100 arcsec is listed in CCDM, probably
optical. 

\item \object{HD 30649}: 
CCDM lists a  companion at 3.37 arcsec = 101 AU, but probably this is not the cause of the 
linear trend reported by \citet{2002ApJS..141..503N}.

\item \object{HD 31412}: 
additional common proper motion companion \citep[CNS3, ][]{2007AJ....133..889L} at 22 arcsec = 792 AU; 
it should not be responsible for the observed RV trend.

\item \object{HD 35956}:
triple system. The primary (the star in the UD sample)
is a spectroscopic and astrometric binary \citep{2002ApJ...568..352V}.
An additional companion ($M=0.44~M_{\odot}$) is at a projected separation
of 99 arcsec = 2860 AU.

\item \object{HD 38529}: it hosts
two low-mass companions  with projected masses 
of 0.78 $M_J$ for the inner companion and 12.70 $M_J$ for the outer companion.
This star shows an astrometric motion (it is marked with G-Flag in the 
Hipparcos Catalogue) which is probably due to the presence of the outer 
companion (HD 38529 C) for which \citet{2006A&A...449..699R} derived a 
mass of $37^{+36}_{-19}M_J$, clearly into the brown dwarf regime.
%If the orbit in this system are coplanar (but this assumption might not 
%be valid, because \citet{2002AAS...201.2404K} showed that this system 
%is dynamically stable even for high mutual inclinations) the derived 
%mass of 2.3 $M_J$ for HD 38529.
A stellar companion at very wide separation is also present.
The classification of this star as a two-planet host with a wide 
stellar companion or a single-planet host with a brown dwarf and another
 wide companion is ambiguous \citep{DB06}.

\item \object{HD 43587}: triple system. The primary (in the UD sample) is
 a spectroscopic binary \citep{2002ApJ...568..352V}.
 An additional  component ($M \sim 0.30~M_{\odot}$, $\rho=
 95$ arcsec =  2860 AU was shown to be physically associated 
 \citep{1991A&A...248..485D}. 
And other 3 faint companions are listed, probably all optical (NLTT);

\item \object{HD 53705}- \object{HD 53706}: another distant companion (K5, mass 0.69)
at 185 arcsec = 3000 AU.

\item \object{HD 65907}: triple system. 
The secondary,  at a projected separation of 60 arcsec
from the primary,  is itself a close
visual binary (projected separation 37 AU). Individual masses
0.63 and 0.35 $M_{\odot}$.

\item \object{HD 72760}:
\citet{metchev} reports a companion of 0.13 $M_{\odot}$ at a separation 
of 0.96 arcsec = 21 AU, which is probably responsible of the astrometric trend found by \citet{2005AJ....129.2420M}.

\item \object{HD 77407}: 
0.30 $M_{\odot}$ companion at 1.6 arcsec = 50 AU separation imaged by Calar Alto 
Adaptive optic system ALFA and confirmed as physically bound to 
the primary by a multi-epoch, high-resolution spectrum \citep{2004A&A...417.1031M} that shows a long-term 
radial velocity trend for \object{HD 77407 A}.  
The companion was also confirmed by \citet{metchev}.

\item \object{HD 86728}: the M dwarf companion is overluminous with an high
activity level. \citet{2000MNRAS.311..385G} suggest it is itself a 
close binary.

\item \object{HD 90839}:
CPM pair. There is a third companion, HD 89862, but it is not 
physical \citep[see][]{1991adc..rept.....G,1991A&A...248..485D}.

\item \object{HD 92222 A}: 
it is not included in the Hipparcos Catalogue. For this reason the mass of this star is not listed in VF05. 
We derived a photometric distance assuming that both components are on the main sequence; and by using the isochrones by 
\citet{2002A&A...391..195G} we found $d=90$ pc and $M_A$= 1.09 $M_{\odot}$, $M_B$ = 1.05 $M_{\odot}$.

\item \object{HIP 52940} - \object{HIP 52942}: triple system.
The primary is a spectroscopic binary discovered by \citet{2002ApJS..141..503N}.
Both components are in the UD sample and both masses are derived according with \citet{1997AJ....113.2246R} 
and \citet{2000A&A...364..217D} mass-luminosity calibrations, 
because VF05 do not list any mass value for these objects.

\item \object{HD 97334}: a close pair of brown dwarfs (separation 1.5 AU, total mass = 0.05 $M_{\odot}$)
is at a projected separation of about 2000 AU 
from the primary \citep{2005AJ....129.2849B}.

\item \object{HD 99492 B}: planet outside the UD limits.

\item \object{HD 101177}: triple system. The secondary is a spectroscopic binary
(P=23 d, masses = 0.74 and 0.31 $M_{\odot}$).
Component C is optical as the other two companions listed most likely are (NLTT).

\item \object{HD 114729}: planet outside the UD limits.

\item \object{HD 120136} = $\tau$ Boo:
the orbital solution by  \citet{1994AJ....107..306H} is very preliminary. 
Another L dwarf companion candidate 
at  42 arcsec = 664 AU has been reported by \citet{2006MNRAS.368.1281P}.
The physical association has yet to be confirmed.

\item \object{HD 120780}: 
CCDM lists a companion at 6 arcsec = 98 AU (M=0.53 $M_{\odot}$), but probably this is not 
responsible for the astrometric signature reported 
by \citet{2005AJ....129.2420M}.

\item \object{HD 128620}-\object{HD128621} ($\alpha$ Cen): Triple system: 
another low mass companion (Proxima Cen) at very wide separation (10000 AU).

\item \object{HD 131156}=$\xi$ Boo:
visual binary ($\rho = 4.9$ arcsec = 33 AU, $P=152$ yr) \citep[see][]{1983PUSNO..24g...1W}.
The visual companion explains the long-term radial velocity trend 
with a significant curvature detected by \citet{2006AJ....132..177W}.
There are two additional components listed in WDS: C ($V$=12.6, 
$\rho= 66.7$ arcsec = 447 AU, from CCDM), which is 
optical \citep{1991A&A...248..485D}, and D ($V=9.6$,  $\rho= 49$ arcsec = 328.3 AU).

\item \object{HD 131977}: quadruple system.
The secondary (\object{HD 131976}) is itself a binary with $a=0.9$ AU and individual
masses 0.57 and 0.38 $M_{\odot}$. The fourth component is the brown
dwarf \object{GL 570 B} at a projected separation of 1500 AU \citep{2000ApJ...531L..57B}.

\item \object{HD 137778}:
wide visual pair with a similar component. Only the secondary (\object{HD 137778}) is
included in the UD sample. The primary (\object{HD 137763}) is a spectroscopic
binary with extreme eccentricity \citep[$e=0.975$;][]{1992A&A...254L..13D}.

\item \object{HD 139323}: hierarchical triple system.
The companion \object{HD 139341} is a visual binary with $a=18$ AU and individual
masses of 0.74 and 0.81 $M_{\odot}$.

\item \object{HD 145958}: another companion candidate has been reported
at 0.2 arcsec = 5 AU by \citet{1997A&AS..124...75T} but it needs confirmation (no RV
variations observed).
Preliminary binary orbit of the wide pair in WDS (grade 4).

\item \object{HD 146362}: member of a hierarchical multiple system.
 The star included in the UD sample is orbiting the nearly equal-mass
 double-lined spectroscopic binary \object{HD 146361} (period 1.1 days, individual
 masses 1.10 and 1.09 $M_{\odot}$). Preliminary orbit of the wide pair in
 WDS.
 Another CMP companion, a faint M3V dwarf, is at 633 arcsec = 14000 AU.

\item \object{HD 161797} ($\mu$ Herculis): 
astrometric orbit by \citet{1994AJ....108.2338H} (period 65 years).
The companion was also identified using adaptive-optics imaging 
by \citet{2001AJ....121.3254T} and \citet{2002ApJ...572L.165D}.
Radial velocity monitoring revealed a long term trend with significant
curvature \citep{2002ApJS..141..503N,2006AJ....132..177W}.
The star is also listed as a $\Delta \mu$ binary 
in \citet{2005AJ....129.2420M}.
Another companion, \object{$\mu^2$ Her B}, lies at 34 arcsec = 285 AU
and is itself a visual binary with period 43.2 yr (WDS).

\item \object{HD 167215}:
CCDM lists a companion with $\rho = 54^{''}.8$ (M=1.05 $M_{\odot}$), but is not expected to be 
responsible of the astrometric signature reported by \citet{2005AJ....129.2420M}.

\item \object{HD 169822}: triple system. The star included in the UD sample is a
spectroscopic binary detected during the Keck survey. 
\citet{2002ApJ...568..352V} derived a combined spectroscopic and
astrometric solution ($M=0.30~M_{\odot}$, $a=0.84$ AU). 
\object{HD 169889}, at 608.4 arcsec = 16426.8 AU, is a CPM star.
The revised distance of \object{HD 169822} by \citet{2002ApJ...568..352V} 
(32 pc), coupled with the common RVs for the two stars
\citep{2004A&A...418..989N} makes the 
indication of a physical association stronger.

\item \object{HD 178911}: triple system \citep{2000AstL...26..116T}.

\item \object{HD 184860}: triple system. The star included in the UD sample is a
spectroscopic binary \citep{2002ApJ...568..352V}. The projected
mass of the companion is in the brown dwarf range ($32~M_{J}$). 
An additional companion is  at 5.0 arcsec = 151 AU.

\item \object{HD 187691}: three components listed in CCDM/WDS. AC is a CPM pair, B is optical, 
C is physical \citep{1991A&A...248..485D}.

\item \object{HD 190067}:  
\citet{2001AJ....121.3254T} listed a companion at 2.86 arcsec = 55.19 AU, confirmed 
by \citet{2002AJ....124.1127C} that reported $m  \sim 0.08-0.10 M_{\odot}$. 

\item \object{HD 190360}:  Planet outside the UD limits.

\item \object{HD 190406} (15 Sge):
A brown dwarf companion ($\rho= 0.79$ arcsec = 14 AU)  was found 
with high-resolution imaging made using adaptive optics at the Gemini-North 
and Keck telescopes \citep{2002ApJ...571..519L}. The primary shows a long-term radial velocity trend 
that confirms that \object{HD 190406 B} is physical, with a minimum mass of 
M = 48 $M_J$.

\item \object{HD 196885}: 
\citet{2006A&A...456.1165C} identified at about 0.7 arcsec 
a relatively bright companion candidate: \object{HD 196885 B}, 
that is likely to be a late K-dwarf, with a mass of 0.6 $M_{\odot}$, 
orbiting \object{HD 196885 A} at a projected physical distance of 25 AU.
\object{HD 196885} was also classified as a $\Delta \mu$ binary by 
\citet{2005AJ....129.2420M}. The astrometric signature is
probably due to \object{HD 196885 B}.
A planetary companion was also claimed from the  Lick RV 
Survey \footnote{http//exoplanets.org/esp/hd196885/hd196885.shtml}, 
but it is not confirmed in the recent work by \citet{2006ApJ...646..505B}.
The companion \object{BD+104351 B}, listed in WDS and CCDM, is probably optical.

\item \object{HD 197076}: 
the companion listed as B in WDS and CCDM ($m_v$=11.6 and $\rho=94$ arcsec = 1974 AU) 
is optical, but the really bound companion is component 
C ($\rho= 125$ arcsec = 2625 AU) \citep{1991A&A...248..485D}.

\item \object{HD 198387}: 
CCDM lists a companion with $\rho = 12.2$ = 511 AU, but probably this is not responsible 
for the astrometric signature reported by \citet{2005AJ....129.2420M}.

\item \object{HD 215578}: the mass of this star is not listed in VF05, so we derive the value listed in Table 8 according 
      with \citet{1997AJ....113.2246R} and \citet{2000A&A...364..217D} mass-luminosity calibration. 

\item \object{HD 219542}: the masses of the components of this star were taken from \citet{2004A&A...420..683D}, 
as there  is some confusion in the identification of the components in the photometry used by VF05 
to derive stellar masses (for one of the components, the joint A+B magnitude is used).

\item \object{HD 219834}: hierarchical triple system. The primary is
a spectroscopic binary (period=6.2 yr, individual masses of 0.90 and 0.04 $M_{\odot}$).
The secondary is the star included in the UD sample and the mass was apparently derived in VF05 from the magnitude of the 
much brighter primary. We assumed the mass  given in the on-line version of the MSC \citep{1997A&AS..124...75T}.
\end{itemize}

\subsection{Unconfirmed binaries}
\label{app:unconfirmed}
\begin{itemize}
\item \object{HD 23249} ($\delta$ Eri): belongs to the group of nearest stars and is classified as a weakly 
active and X-ray soft source \citep{1998A&AS..132..155H}.
\citet{1983IBVS.2259....1F} tried to detect a periodic variation in the 
photometric data and suggest that $\delta$ Eri could be classified as an 
RS CVn star: that is, a F-G binary star having a period $<$ 14 days, with chromospheric 
activity and with a period of rotation synchronized with its orbital 
period \citep{1984STIN...8513704L}, giving the star high rotational 
velocity inducing strong activity.
This contrasts with the low level of activity detected 
by \citet{2005A&A...436..253T} and the lack of radial
velocity variations, making this classification doubtful.

\item \object{HD 52265}, \object{HD 154857}, \object{HD 179949}: first epoch observations 
by \citet{2006A&A...456.1165C} revealed companion candidates, 
but the physical association with the planet hosts has not yet been confirmed;

\item \object{HD 102158}: the presence of a common proper motion companion at
1175.2 arcsec was proposed by \citet{2007AJ....133..889L}. The
corresponding projected separation (61000 AU) is much larger than any
other companion in Table 8, so it is not considered here.

\item \object{HD 107213}: \citet{2007AJ....133..889L} propose that this star form
a wide (projected separation 546 arcsec) common proper motion pair with
\object{BD+28 2103} = \object{HIP 60061}. However, the discrepant RVs for the two stars
\citep{2002AJ....124.1144L,2004A&A...418..989N}
argue against a physical association.

\item \object{HD 117176}: an L dwarf companion candidate 
at 848 AU has been reported by \citet{2006MNRAS.368.1281P}.
The physical association has yet to be confirmed.

\item \object{HD 168443}: as for \object{HD 38529}, there is an ambiguity about the physical
classification of the companion as a massive planet or brown 
dwarf \citep[see][]{DB06}.

\item \object{HD 217107}: the presence of the companion listed in WDS is not confirmed 
by recent adaptive optics searches.
See \citet{DB06} for discussion and references.

\end{itemize}

\end{appendix}

\newpage

\end{document}